\documentclass[letterpaper, 11pt]{article}

\usepackage{graphicx}
\usepackage[colorlinks=true,
            linkcolor=blue,
            urlcolor =blue,
            citecolor=blue,
            anchorcolor=blue]{hyperref}
\usepackage{amsmath}
\usepackage{arydshln}
\usepackage{booktabs}
\usepackage{caption}
\usepackage{changepage}
\usepackage{comment}
\usepackage[margin=1.0in, showframe=false]{geometry}
\usepackage[utf8]{inputenc}
\usepackage{lscape}
\usepackage{multirow}
\usepackage{natbib}  % for citation hyperlinks
\usepackage{setspace}
\usepackage{titling}
\usepackage[flushleft]{threeparttable}
\usepackage{xcolor}
\hypersetup{colorlinks,citecolor=blue}  % set citation hyperlinks blue
\title{\textbf{Voting From Jail}\thanks{We thank Roger Pharr and Andrea Wang for their contributions to data collection and matching. We thank Ted Enamorado and Ariel White for their helpful insights. This project was supported by Arnold Ventures and the Chan Zuckerberg Initiative.}\\\vspace{5mm}WORKING PAPER}
\author{Anna Harvey\thanks{Department of Politics, Public Safety Lab, New York University, \href{mailto:anna.harvey@nyu.edu}{anna.harvey@nyu.edu}}  \and Orion Taylor\thanks{Public Safety Lab, New York University,   \href{mailto:ojt212@nyu.edu}{ojt212@nyu.edu}}}

\begin{document}
\begin{titlingpage}
\maketitle
\begin{abstract} \noindent
We leverage new data on daily individual-level jail records and exploit the timing of incarceration to estimate the causal effects of jail incarceration on voting from jail in 2020. We find that registered voters booked into county jails for the full duration of 2020 voting days were on average 46\% less likely to vote in 2020, relative to registered voters booked into the same jails within 7 -- 42 days after Election Day. The estimated negative effect of incarceration on voting from jail was much larger for Black registered voters, who were 78\% less likely to vote in 2020 if booked into county jails for the full duration of 2020 voting days, relative to Black registered voters booked into the same jails just after Election Day. Placebo tests indicate no effects of 2020 jail incarceration on the 2012 or 2016 turnout of registered voters. We find inconsistent effects of jail incarceration on voter registration in 2020, and effect sizes of comparable magnitude for turnout unconditional on registration status. Our findings reveal the pressing need to enable voting-eligible incarcerated individuals to exercise their constitutional right to vote, and to address troubling racial disparities in the effect of jail incarceration on the exercise of the right to vote.
\end{abstract}
\end{titlingpage}

\begin{onehalfspacing}

\section{Introduction}

In every state, otherwise voting-eligible individuals incarcerated pretrial or to serve a misdemeanor sentence remain legally entitled to vote while incarcerated \citep{SP_20, SP_July22}. The vast majority of the approximately 650,000 individuals incarcerated in county jails on any given day are being held pretrial or to serve misdemeanor sentences \citep{whole_pie}. Anecdotal evidence suggests, however, that many of those incarcerated in county jails are not being given adequate opportunities to exercise their right to vote \citep{SP_20}. Concerns have also been raised about possible racial disparities in ballot access for those incarcerated in county jails \citep{SP_July22}. Yet we lack reliable causal estimates of the impacts of being incarcerated in a county jail on the exercise of the right to vote, and of any racial heterogeneity in those impacts.

Some recent papers have sought to estimate the impacts of prison and jail incarceration on individuals' post-release voting behavior \citep{gerber, white2019a, mcdonough}. \cite{white2022} describe the extent of voting from prison in Maine and Vermont---the two states permitting individuals serving felony sentences to vote from prison---but do not estimate the causal impacts of prison incarceration on voting from prison.  

We leverage new data on daily individual-level jail records from the Jail Data Initiative and exploit the timing of incarceration to estimate the causal effects of being incarcerated in a county jail during 2020 general election voting days on the probability of voting. We probabilistically match 944,985 individual-level booking records from 936 jail rosters with 195,655,326 voter records from 42 corresponding statewide files for a period of 180 days centered on Election Day (November 3, 2020). We identify individuals whose periods of jail incarceration began during 2020 voting days in their state (either mail-in or in-person), or within windows ranging between 7 and 42 days after Election Day. Our primary analyses are conducted within the sample of jailed individuals who match to voter records with match probability $p>0.75$; we also replicate analyses for the sample with match probability $p>0.95$. We interpret matches as indicating registered voters who were incarcerated in county jails during our time period of interest. We source information about jailed registered voters from both jail and voter records. For post-Election Day control group windows ranging between 7 and 42 days, we conduct a series of balance tests on individual-level and booking-level characteristics, including age, race, gender, partisanship, and number and kind of booking charges, to identify the pre-Election Day treatment group windows within which individuals booked into county jails during 2020 voting days (either mail-in or in-person) were observably indistinguishable (joint F-test p-value $>$ 0.10) from individuals booked into the same jails during post-Election Day control group windows. For records with match probability $p>0.75$, the balanced samples range in size from 57,821 (28-day control group window) to 103,091 (42-day control group window). 

\begin{comment}
Using a threshold match probability of 0.75, we identify 331,703 (?) matches across jail and voter records, which we interpret as a proxy for registered voters incarcerated in county jails during this period.
\end{comment}

Our identification strategy rests on the assumption that individuals booked into the same jail within a narrow temporal window, who are indistinguishable on observed characteristics, are also indistinguishable on unobserved characteristics. If this assumption is correct, we can use the 2020 turnout of individuals booked into county jails just after the last 2020 voting day as a valid counterfactual for the turnout we likely would have observed among individuals booked into the same jails during 2020 voting days, had they not been incarcerated during voting days.

Turnout among registered voters in 2020 is estimated by the U.S. Census Bureau to have been 91.9\%.\footnote{\url{https://www.census.gov/content/dam/Census/library/publications/2022/demo/p20-585.pdf}} Registered voters booked into county jails just after the last 2020 voting day had much lower average turnout in 2020, relative to the general population, underscoring the importance of identifying an appropriate counterfactual turnout rate for those jailed during 2020 voting days. For the sample with match probability $p>0.75$, turnout among registered voters who were booked into county jails within 7 -- 42 days after 2020 Election Day ranged between 34.9\% and 35.7\% in our samples, or between 56\% and 62\% less than the registered voter turnout observed in the U.S. population. With a less appropriate control group, we might mistakenly attribute some of the decreased turnout among those jailed prior to the election to the effects of jail incarceration, rather than to a lower propensity to vote.

Within our balanced samples, we estimate the impacts of (1) being booked into a county jail during 2020 voting days (either mail-in or in-person) for any length of time and (2) being booked into a county jail during 2020 voting days for some proportion of voting days on the probability that a registered voter voted in the 2020 election. In models that include all individual- and booking-level covariates along with jail and week fixed effects, we find that registered voters who were booked into county jails during 2020 voting days for any length of time were on average 3.0 -- 3.3 percentage points or 8.6\% -- 9.3\% less likely to have voted in 2020, relative to registered voters who were booked into the same jails within 7 -- 42 days after Election Day (avg. effect magnitude = 9.0\%; p $<$ 0.01).

The impacts of incarceration on voting from jail in 2020 increased with the duration of incarceration. Including all individual- and booking-level covariates along with jail and week fixed effects, registered voters who were booked into county jails during 2020 voting days for the full span of voting days in their state were on average 13.7 -- 22.9 percentage points or 38.4\% -- 65.2\% less likely to have voted in 2020, relative to registered voters who were booked into the same jails within 7 -- 42 days after Election Day (avg. effect magnitude = 46.1\%; p $<$ 0.01). These estimates imply that registered voters who were booked into county jails during 2020 voting days for the full span of voting in their state voted from jail at rates ranging between 12.2\% and 22.0\%. In a series of placebo tests, we find no consistent effects of the timing of individuals' 2020 bookings on their probabilities of having voted in the 2012 or 2016 elections. All estimates are of comparable magnitude in the sample of linked records with match probability $p>0.95$. 

We also explore racial heterogeneity in the effects of incarceration on voting from jail in 2020, finding that the effect of jail incarceration was significantly larger for Black registered voters, relative to white registered voters. White registered voters booked into county jails just after Election Day 2020 had turnout rates ranging between 37.7\% and 38.3\%, considerably higher than the turnout rates for Black registered voters booked into county jails just after Election Day 2020, which ranged between 27.8\% and 28.2\%. Jail incarceration also had less of a negative effect on turnout for white registered voters. Including all individual- and booking-level covariates along with jail and week fixed effects, white registered voters who were booked into county jails during 2020 voting days for the full set of voting days were on average 12.5 -- 22.8 percentage points or 32.6\% -- 60.2\% less likely to have voted in 2020, relative to white registered voters who were booked into the same jails within 7 -- 42 days after Election Day (avg. effect magnitude = 39.8\%; p $<$ 0.01). These estimates imply that white registered voters booked into county jails during 2020 voting days for the full set of voting days voted from jail at rates ranging between 15.1\% -- 25.8\%. The estimated negative effect of incarceration on voting from jail was an additional 5.0 -- 11.8 percentage points or 38.2\% -- 55.0\% larger for Black registered voters (avg. effect magnitude = 45.9\%; p $<$ 0.05). The total estimated effect of jail incarceration for Black registered voters booked into county jails during 2020 voting days for the full span of voting days was a 17.9 -- 34.6 percentage point or 61.3\% -- 121.8\%  decrease in turnout, relative to Black registered voters who were booked into the same jails within 7 -- 42 days after Election Day (avg. effect magnitude = 77.9\%). These estimates imply that Black registered voters booked into county jails during 2020 voting days for the full set of voting days voted from jail at rates ranging between 0.0\% and 11.3\%. Estimates of racially heterogeneous effects are of comparable magnitude when we restrict the sample to states that report voter race (eliminating the records for which race is predicted by L2).

We also explore the effects of jail incarceration on voter registration behavior, and on voting behavior unconditional on registration status. Jail incarceration in 2020 had inconsistent effects on voter registration. The effects of jail incarceration on turnout in 2020 unconditional on voter registration are of similar magnitude to the effects estimated from the sample of registered voters only. 

The nature of our data prevent us from exploring the precise mechanisms generating both the overall decreases in turnout for individuals booked into county jails during 2020 voting days, and the substantially larger decreases for Black individuals. Nevertheless, at a minimum our findings reveal a pressing need for states and counties to take steps to ensure that voting-eligible incarcerated individuals are given adequate opportunities to exercise their constitutional right to vote, and to address troubling racial disparities in the exercise of that right.

\section{Incarceration and Voting}

Whether and by how much being incarcerated in a county jail interferes with an individual's exercise of the right to vote, and whether there are any racial disparities in the effects of incarceration on voting from jail, are empirical questions. We lack systematic evidence on these questions.

%Using survey data and conditioning on observable characteristics, \cite{weaverlerman2010} and \cite{lerman2014} reported negative associations between self-reported incarceration spells and self-reported post-release voting. However, self-reports of both post-release voting and incarceration spells may be affected by systematic measurement error \citep{gerber}.

Some researchers have sought to estimate the post-release impacts of periods of prison and jail incarceration on individuals' exercise of their right to vote \citep{gerber, white2019a, mcdonough}. \cite{gerber} used state prison and sentencing records to investigate the impacts of prior periods of prison incarceration on post-release voting in the 2012 election, finding few differences in turnout between those incarcerated and released from state prison between 2008 and 2012 and either observably comparable individuals in the Pennsylvania voter files, or observably comparable convicted defendants who received sentences other than prison incarceration. \cite{gerber} also found few differences in post-release turnout between those convicted defendants sentenced to jail incarceration on their most serious charge and those receiving non-carceral sentences.

\cite{white2019a} addressed potential bias arising from selection into jail incarceration by estimating the effects of a sentence to jail incarceration on post-release 2012 turnout using a design based on the as-if random assignment of first-time misdemeanor defendants to judges in Harris County, Texas. \cite{white2019a} found a 13 percentage point average decrease in post-release 2012 turnout among marginal Black misdemeanor defendants sentenced between 2008 and 2012, and negligible effects for marginal white defendants. 

\cite{mcdonough} used a similar design to estimate impacts from pretrial jail incarceration between 2008 and 2012 in Philadelphia County on post-release 2012 turnout, leveraging the assignment of defendants to bail judges to attempt to identify causal effects. \cite{mcdonough} found a 39\% average reduction in post-release voter turnout among marginal defendants incarcerated pretrial before the 2012 election, with effects appearing in the year prior to the election and with larger effects for Black defendants.\footnote{The assignment of bail judges to cases in Philadelphia County during this period is not independent of case and defendant characteristics (p $<$ 0.001), raising the possibility that non-randomly assigned case and defendant characteristics may account for the larger estimated reduction in post-release turnout, relative to \cite{white2019a}.} 

 Incarceration may affect turnout not only after a defendant is released, but also, for those defendants incarcerated during voting days, during periods of incarceration \citep{SP_20, SP_July22}. \cite{white2022} merged state prison records with voter file records in Maine and Vermont---the two states in which individuals incarcerated for felony convictions are legally entitled to vote while incarcerated---finding that 8\% and 6\% of those incarcerated in state prison voted while incarcerated in 2018 in Vermont and Maine, respectively. \cite{white2022} did not attempt to estimate the causal impacts of incarceration on the incidence of voting from prison. \cite{mcdonough} attempted to distinguish the incapacitation effect of jail incarceration on Election Day 2012 from the post-release effects of prior periods of jail incarceration, but their design and data did not allow them to cleanly identify an incapacitation effect. 
 
 To date we lack any reliable causal estimates of whether and by how much being incarcerated in a county jail interferes with an individual's exercise of the right to vote. Individuals incarcerated in county jails are, with few exceptions, not rendered ineligible to vote because of their jail incarceration.\footnote{The vast majority of those held in county jails are being held pretrial or for the purpose of serving a sentence following a misdemeanor conviction, neither of which is disenfranchising in any state. Like individuals not incarcerated in county jails, those incarcerated in county jails may be ineligible to register and vote for reasons unrelated to their current incarceration, including citizenship status and prior felony convictions. For more information on enfranchisement status within incarcerated populations, see \cite{SP_20}.} Anecdotal reports suggest that county jails nonetheless often fail to provide incarcerated individuals with the ballot access to which they are constitutionally entitled \citep{SP_20,SP_July22}. However, we lack reliable estimates of the magnitude of the impacts of jail incarceration on voting from jail. We also lack any systematic estimates of racial disparities in these impacts. The lack of rigorous evidence about these impacts impoverishes policy debates.
 
 We probabilistically match daily incarceration records from county jails in the United States to voting records from the 2020 general election, estimating the impacts of jail incarceration on turnout for those whose periods of jail incarceration began during 2020 voting days, relative to those whose periods of jail incarceration began just after the last 2020 voting day. 

\section{Data}

\subsection{Jail Incarceration Data} \label{data_jdi}

We source daily individual-level jail incarceration data from the Jail Data Initiative (JDI), a project of New York University's Public Safety Lab.\footnote{\url{https://jaildatainitiative.org/}.} Daily JDI jail records are scraped from county, sheriff, and corrections websites posting public rosters of all incarcerated individuals on (at least) a daily basis. By Fall 2020, the JDI project was scraping daily jail rosters from more than 1,100 county jails. County jails are not included in the JDI data if they do not post online jail rosters (e.g., all jails in the state of Alaska), or if their states have unified correctional systems (e.g., Connecticut and Hawaii).

Daily jail roster records are converted into booking records by identifying unique individuals and unique periods of incarceration within each facility. Unique individuals are identified using a variety of fields including booking number, person identification number, name, and date of birth, depending on field availability. Unique periods of incarceration or booking records are identified using the dates on which unique individuals enter and exit jail rosters (unique individuals may have multiple booking records in the same county). Demographic and bail/bond fields are standardized across facilities. Charge descriptions contained in text fields are classified into charge categories using the Text-Based Offense Classification algorithm developed by the Criminal Justice Administrative Records System at the University of Michigan \citep{cjars}. 

The JDI bookings database contains daily jail incarceration data unprecedented in granularity and national scope. While the Bureau of Justice Statistics produces a 5-year Census of Jails---most recently in 2019---and an intervening Annual Survey of Jails, these data are only based on single-day point-in-time snapshots of jail populations \citep{bjs}.

For our main analyses of the effects of jail incarceration on the turnout of registered voters, we include as covariates JDI data on the length of each booking in days, number of charges associated with each booking, and indicators for whether the most serious CJARS charge category associated with each booking was violent, property, drug, public order, or DUI, with criminal traffic as the excluded charge category.\footnote{The CJARS Text-Based Offense Classification algorithm uses the Uniform Crime Classification Standard (UCCS) \citep{cjars}. We determine most serious charge using a hierarchical ranking of top-level UCCS codes based on the following order of severity, from most to least serious: violent, property, drug, public order, DUI, criminal traffic.} For our analyses of the effects of jail incarceration on voter registration status, and on turnout unconditional on registration status, we also include as covariates JDI data on gender, age, and race. Further information on JDI data collection and standardization is provided in Appendix \ref{app:c}.

\subsection{Voter File Data}

We source voter file data from the L2 voter database licensed to New York University. These data contain voting history and demographic information for registered voters.\footnote{\url{ https://l2-data.com/our-data/}.} L2 data were pulled in early 2021. For our main analyses of the effects of jail incarceration on the turnout of registered voters, we include as covariates L2 data on age, gender, race/ethnicity (both reported and predicted), and party registration/identification (both reported and predicted). 

L2 race and ethnicity categories are both reported directly by states and, where not reported by states, predicted by L2 from other features available in the data. We map L2 race/ethnicity values to non-Hispanic white (European), Black (Likely African-American) and Other (East and South Asian, Hispanic and Portuguese, or Other). We replicate our analyses of racially heterogeneous effects of jail incarceration using only states that directly report race data to L2.

L2 party categories are likewise both reported directly by states and, where not reported by states, predicted by L2 from other features available in the data. We map the L2 party categories to Democratic, Republican, and Non-Partisan or Other (a group that includes Non-Partisan, Other, and 54 other minor party descriptions). Further information on L2 voter data is provided in Appendix \ref{app:c}.

\subsection{Matching Jail Records to Voter File Records}

We initially constructed a sample of 944,985 JDI jail booking records from 936 counties for individuals who were booked into county jails on dates within a +/- 90-day window around Election Day 2020 (August 5, 2020 to February 1, 2021). To make probabilistic record linkage tractable, we created age strata for JDI bookings and blocked on L2 within-state matching pools of +/- 2 years. Where age was missing for JDI booking records, we matched across full-state L2 voter files. We further restricted blocks to L2 surname soundex matches of incarcerated individuals.

Data completeness varies by jail roster. Due to the variety in data quality, we employed a relaxed, Java-implemented modification of the probabilistic matching algorithm used by \cite{mcdonough}, itself an instance of the Fellegi-Sunter/Expectation Maximization linkage model as developed by \cite{enamorado}. In lieu of labeled matches we produced a baseline averaged Jaro-Winkler distance between name components, and considered all scores above a threshold to be matches for parametric initialization. We then vectorized similarities of name, gender, and in-block age as ternary (to account for missingness and variation in reporting of name components), and FIPS codes as binary (within-state). We ran samples from this pool of match-candidate pairs through a standard expectation maximization process \citep{winkler} to stabilize parameters, and then conducted our final linkage, keeping only highest-scoring matches per booked individual and discarding low-probability matches.

We followed a similar strategy of probability match score re-weighting based on first and last name frequencies to that employed by \cite{enamorado} and \cite{mcdonough}, and then further thresholded matched results. We also removed any rows indicating multiple voter matches per booked individual, where multiple matched bookings exhibited temporal overlap, or where booking records indicate that individuals were younger than 18 years old. Our primary analysis sample includes linked records that exceed a match probability of 0.75; we also replicate analyses with linked records that exceed a match probability of 0.95.  

\subsection{2020 Voting Days}

In response to the COVID-19 pandemic, many states enacted emergency measures that impacted voting days and methods during the 2020 general election. Some states suspended requirements for requesting absentee mail ballots, instead permitting no-excuse absentee ballot requests; others automatically sent out mail-in absentee ballots to registered voters for the first time, or extended early in-person voting days \citep{brennan_center}. There was a 105\% increase in voting by mail from the 2016 general election to the 2020 general election, and a 37\% increase in early in-person voting \citep{census_voting}.

Mail-in ballot and early in-person voting days varied both across and within states. For each state we record the earliest of the two dates on which voters could begin (1) returning the mail-in ballots mailed out by county boards of elections, or (2) voting in person. Earliest mail-in ballot dates range from September 4, 2020 (60 days prior to Election Day) in North Carolina to October 16, 2020 (19 days prior to Election Day) in Washington state. Earliest in-person voting dates range from September 14, 2020 (50 days prior to Election Day) in Pennsylvania to Election Day itself in several states \citep{ncsl, cnbc}. By taking the earlier of these two dates per state, we ensure our treatment windows cover the complete legal voting periods of each state.

\section{Analysis}

Individuals incarcerated in jails or prisons likely differ along a number of dimensions, relative to unincarcerated individuals. These differences may be related to individuals' propensity to vote. Of particular concern, individuals at high risk for incarceration may have low propensity to vote, leading researchers to overestimate the impacts of incarceration on voting. 

\cite{gerber} sought to address selection into incarceration by controlling for observed characteristics correlated with both risk of incarceration and propensity to vote. However, this strategy does not address the possible presence of unobserved characteristics related to both risk of incarceration and propensity to vote. \cite{white2019b} and \cite{mcdonough} addressed selection into jail incarceration by leveraging the potentially as-if random assignment of cases to trial and bail judges. However, this strategy does not permit clean identification of the effects of incarceration on voting from jail, as whether an individual remains incarcerated during legal voting periods is affected by factors other than a judge's bail determination or sentencing decision \citep{mcdonough}.

We leverage the timing of jail bookings to address selection into jail incarceration. Individuals booked into county jails during legal voting days in 2020 are likely very similar to individuals booked into those same jails just after the last legal voting day (November 3, 2020 in all states). In our design, incarcerated individuals are effectively randomized into treatment and control groups based on the timing of their jail incarceration, with treated individuals booked into county jails during legal voting days, and control individuals booked into those same jails just after Election Day. This identification strategy has been successfully deployed in a number of recent papers \citep{white2019a, MMS, doleac}. Following this literature, we report balance tests to confirm the similarity of treatment and control groups.

Incarceration during 2020 voting days likely had a larger effect on the voting behavior of registered voters than on the registration (and possibly voting) behavior of unregistered voters. We initially restrict our attention to estimating the effect of jail incarceration on the voting behavior of registered voters booked into county jails during our period of interest, using a match probability of $p>0.75$ as indicating a jailed registered voter. We replicate analyses for records that have a match probability of $p>0.95$. We then estimate the effects of jail incarceration on both registration and voting behavior unconditional on registration status.

\subsection{Balance Tests}

We expect that registered voters booked into county jails during 2020 voting days will have similar individual- and booking-level characteristics as those booked into the same jails just after the close of legal voting (November 3, 2020 in all states). However, in 2020 voting in some states commenced as early as 60 days prior to Election Day (mail-in voting in North Carolina). Registered voters booked into county jails 60 days prior to November 3, 2020 may not have been comparable to registered voters booked into those same jails just after November 3, 2020.

To identify temporal windows within which individual- and booking-level characteristics are balanced across treatment and control groups, we estimated a series of balance tests for windows of varying lengths including both 2020 voting days and post-Election control days. We assessed balance on a variety of individual- and booking-level characteristics, including age (from L2, in years), race/ethnicity (from L2, indicators for Black and non-Hispanic white), party registration/identification (from L2, indicators for Democratic and Republican registration), gender (from L2, indicator for male), number of charges (from JDI), and indicators for whether the top charge against the booked individual was a DUI, drug, property, public order, or violent charge (from JDI). Samples include linked records (registered voters booked into county jails during the specified time periods) with match probability $p>0.75$. We excluded observations missing data on any covariates, or where voting records indicated that an individual did not register to vote until after November 3, 2020. All balance tests included jail-level fixed effects and clustered standard errors on jail facilities. 

\begin{center}
\begin{figure}[h]
\captionof{figure}{Balance Tests (Joint F-test P-values)}\label{fig:pvals_0.75}
\centering
\includegraphics[width=\textwidth]{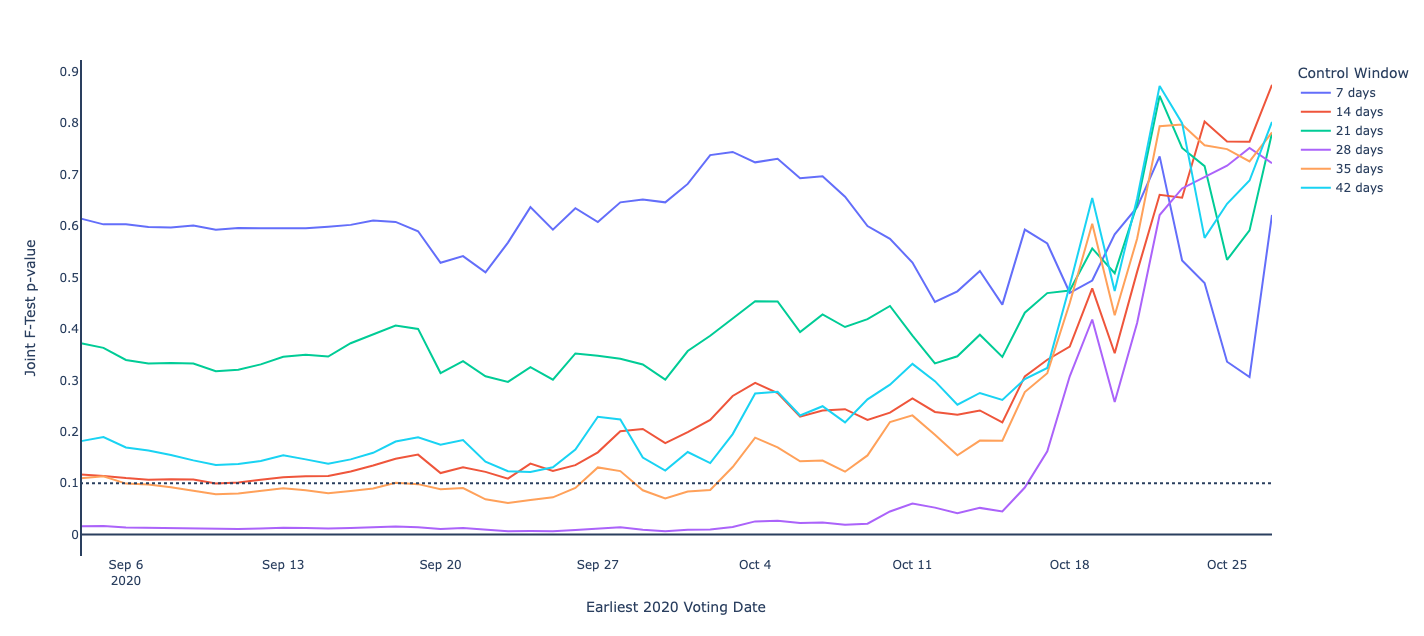}
\begin{tablenotes}
\small
\item \textbf{Note:} This figure reports estimates of balance on individual- and booking-level characteristics for post-Election Day control windows ranging between 7 and 42 days and treatment windows containing legal voting days in 2020, for the sample of matched records (registered voters booked into county jails during this time period) with match probability $p>0.75$. For each post-Election Day control window we initially tested for balance using treatment windows containing the full set of voting days in each state in 2020, sequentially dropping the earliest voting day in the sample and repeating the balance test until we reached a treatment window of 7 days prior to Election Day.  
\end{tablenotes}
\end{figure}
\end{center}

Figure \ref{fig:pvals_0.75} reports estimates of balance on individual- and booking-level characteristics for post-Election Day control windows ranging between 7 and 42 days and treatment windows containing legal voting days in 2020. For each post-Election Day control window we initially tested for balance (joint p-value $>$ 0.10) using treatment windows containing the full set of voting days in each state in 2020. We sequentially dropped the earliest voting day in the sample and repeated the balance test, continuing until we reached a treatment window of 7 days prior to Election Day. Consistent with expectations, treatment group windows closer to Election Day are more consistently balanced with post-Election Day control groups. However, some post-Election Day control groups (7 days, 21 days, and 42 days) are balanced for the full set of 2020 voting days, and stay balanced for all subsets of voting days. Other post-Election Day control groups (14 days, 28 days, and 35 days) are only consistently balanced with subsets of 2020 voting days that are closer to Election Day. 

Table \ref{tab:random_0.75} reports estimates of balance on individual- and booking-level characteristics for post-Election Day control windows ranging between 7 and 42 days and the largest treatment windows for which the joint F-test p-values for that treatment window and for all smaller treatment windows are greater than 0.10. For example, for a 7-day control window, we secure a balanced sample with a 60-day treatment window (p $<$ 0.61). The joint F-test p-values for all treatment windows less than 60 days are also greater than 0.10 for this control group. The treatment group contains bookings beginning on September 4 through November 3 in North Carolina, where absentee ballot mail-out began on September 4; bookings beginning on October 16 through November 3 in Washington state, where both absentee ballot mail-out and early in-person voting began on October 16; and so on. The control window contains bookings beginning on November 4 through November 10. The outcome is whether a booking occurred during a treatment window (1) versus a control window (0). All models include jail-level fixed effects; robust standard errors clustered on jails are reported in parentheses.  

\begin{center}
\begin{figure}[h]
\captionof{figure}{Distribution of Early Ballot Return}\label{fig:early}
\centering
\includegraphics[width=\textwidth]{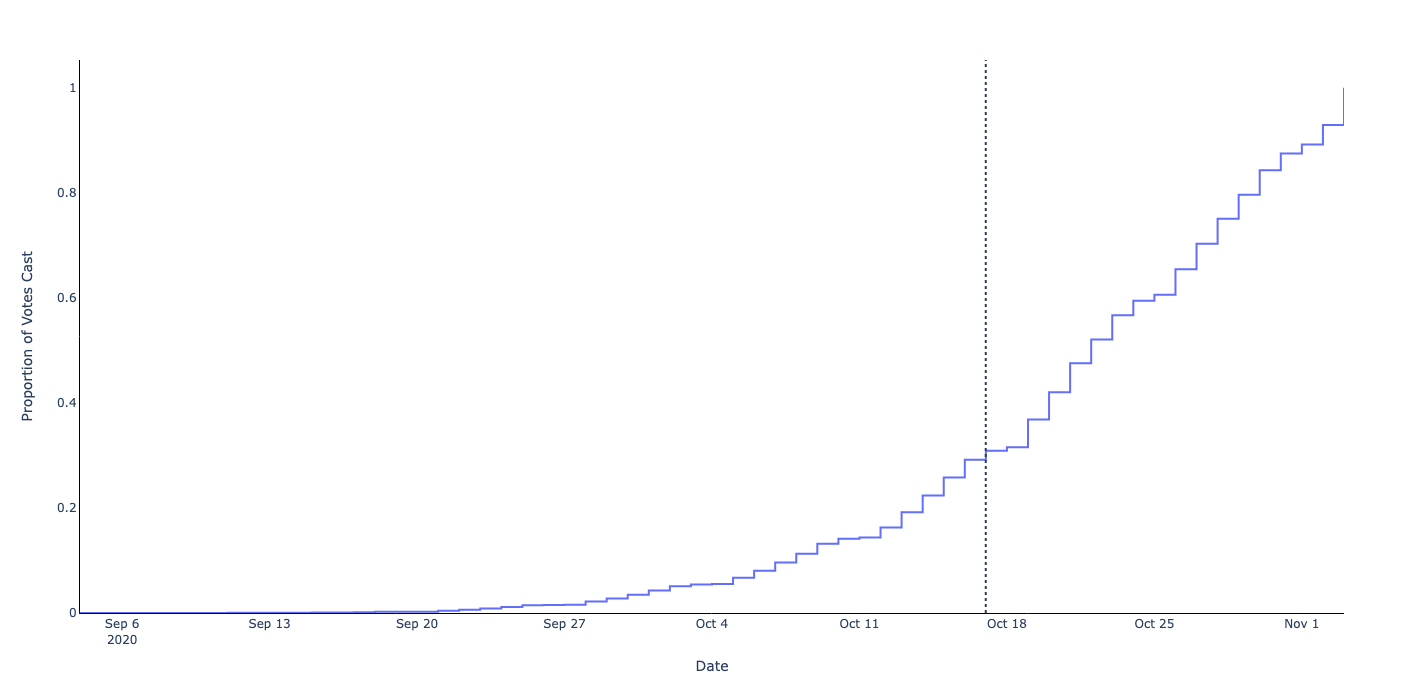}
\begin{tablenotes}
\small
\item \textbf{Note:} This figure reports the cumulative distribution of early ballot return in the L2 voter file data for 2020, where available. The smallest treatment window in the samples reported in Table \ref{tab:random_0.75} begins on October 17, indicated with a vertical dashed line. Approximately 70\% of 2020 ballots were returned after October 17.
\end{tablenotes}
\end{figure}
\end{center}

The L2 data include a ballot return date field indicating early voting. We can use this field to assess the coverage of our treatment windows. Figure \ref{fig:early} reports the cumulative distribution of ballot return over time. The smallest treatment window among the samples in Table \ref{tab:random_0.75} (17 voting days---beginning October 17---with a control group of 28 post-Election days) captures approximately 70\% of ballot return. 

\subsection{Descriptive Statistics}

Table \ref{tab:stats_0.75} reports summary statistics for treatment and control groups for each of the balanced samples identified in Table \ref{tab:random_0.75}. 73.1\% -- 73.7\% of our samples are male. 24.6\% - 27.1\% of our samples are Black, 59.3\% -- 60.7\% are white, and 12.8\% -- 15.1\% are other race. The average age in our sample ranges between 37.2 -- 37.4 years. 19.2\% - 20.4\% of our samples are Republicans, 40.9\% -- 41.9\% are Democrats, and 37.8\% -- 40.0\% are nonpartisan or other party. 

28.3\% -- 29.8\% of those in our samples are incarcerated for public order top charges; 27.4\% -- 29.2\% are incarcerated for violent top charges; 19.4\% -- 20.2\% are incarcerated for property top charges; 12.7\% -- 13.8\% are incarcerated for drug top charges; 6.2\% -- 6.7\% are incarcerated for DUI top charges; 3.2\% -- 3.5\% are incarcerated for criminal traffic top charges. Average number of charges is 2.5 charges across all samples. Average length of jail incarceration ranges from 33.9 -- 37.5 days. The average proportion of 2020 voting days during which individuals are confined ranges between 0.14 -- 0.21.

Turnout among registered voters who were booked into county jails within 7 -- 42 days after 2020 Election Day ranged between 34.9\% and 35.7\% in our samples, or between 56\% and 62\% less than the registered voter turnout observed in the U.S. population.\footnote{\url{https://www.census.gov/content/dam/Census/library/publications/2022/demo/p20-585.pdf}}. As noted earlier, without an appropriate control group, we might mistakenly attribute some of the lower turnout of those incarcerated during 2020 voting days to the effects of jail incarceration, rather than to a lower propensity to vote.

\subsection{Estimating Average Treatment Effects}

Table \ref{tab:turnout_0.75} reports OLS estimates of the effects of (1) being booked into a county jail during 2020 voting days (either mail-in or in-person) for any length of time on a registered voter's probability of voting in the 2020 election, with and without covariates, and (2) being booked into a county jail during 2020 voting days for some proportion of voting days on a registered voter's probability of voting in the 2020 election, with and without covariates. For example, if a registered voter is incarcerated in North Carolina from September 4, 2020 to January 1, 2021, the proportion of voting days confined = 1.0, as the individual is incarcerated during 60 out of all 60 possible voting days in North Carolina. If a registered voter is confined in Washington state from November 3, 2020 to December 10, 2020, the proportion of voting days confined = 0.053, as the individual is incarcerated during 1 out of 19 possible voting days in Washington state. Models are estimated on the samples reported in Table \ref{tab:random_0.75}. All models include jail and week-of-year fixed effects; robust standard errors two-way clustered at the jail and week-of-year level are reported in parentheses. 

As reported in Table \ref{tab:turnout_0.75}, in models that include all individual- and booking-level covariates along with jail and week fixed effects, we find that registered voters who were booked into county jails during 2020 voting days in their state for any length of time were on average 3.0 -- 3.3 percentage points or 8.6\% -- 9.3\% less likely to have voted in 2020, relative to registered voters who were booked into the same jails within 7 -- 42 days after Election Day (avg. effect magnitude = 9.0\%; p $<$ 0.01).

The impacts of incarceration on voting from jail in 2020 increased with the duration of incarceration. As reported in Table \ref{tab:turnout_0.75}, including all individual- and booking-level covariates along with jail and week fixed effects, registered voters who were booked into county jails during 2020 voting days in their state for the full duration of voting days were on average 13.7 - 22.9 percentage points or 38.4\% -- 65.2\% less likely to have voted in 2020, relative to registered voters who were booked into the same jails within 7 -- 42 days after Election Day (avg. effect magnitude = 46.1\%; p $<$ 0.01). 

\subsection{Placebo Tests}

Table \ref{tab:placebo_0.75} reports placebo tests of the effects on 2016 and 2012 voter turnout of a booking occurring during a pre-Election Day window in 2020, relative to a post-Election Day window. The 2016 placebo samples include registered voters in each of our balanced samples who were registered to vote on both Election Day 2020 and Election Day 2016 (November 8, 2016). The 2016 placebo tests report OLS estimates of the effect of a 2020 booking occurring in a pre-Election Day 2020 window on a registered voter's probability of voting in the 2016 election, for the registered voters in each of the balanced samples reported in Table \ref{tab:random_0.75}. The 2012 placebo samples include registered voters in each of our balanced samples who were registered to vote on both Election Day 2020 and Election Day 2012 (November 6, 2012). The 2012 placebo tests report OLS estimates of the effect of a 2020 booking occurring in a pre-Election Day 2020 window on a registered voter's probability of voting in the 2012 election, for the registered voters in each of the balanced samples reported in Table \ref{tab:random_0.75}.

As reported in Table \ref{tab:placebo_0.75}, and as expected, there are no consistent effects of pre-Election Day jail incarceration in 2020 on registered voters' turnout in 2016 or 2012. The effects of 2020 jail incarceration on 2016 turnout are very slightly positive but consistently insignificant. The effects of 2020 jail incarceration on 2012 turnout are very slightly negative but generally insignificant, reaching p $<$ 0.05 in only one balanced sample. By contrast, the effects of 2020 jail incarceration on 2020 turnout among registered voters who were also registered to vote in 2016 and/or 2012 are consistently negative, large, and significant, ranging between 2 -- 4 percentage point or 5.0\% -- 8.8\% decreases (p $<$ 0.01).

\subsection{Racially Disparate Effects of Jail Incarceration}

Concerns have also been raised about possible racial disparities in ballot access for those incarcerated in county jails \citep{SP_20, SP_July22}. We explore racial heterogeneity in the effects of incarceration on voting from jail in 2020 by interacting our two treatment variables (a binary indicator for having been booked into a county jail during 2020 voting days for any length of time, and the proportion of 2020 voting days during which a registered voter booked into a county jail during 2020 voting days was incarcerated) with the indicator for Black registered voters from the L2 voter files. For ease of interpretation we retain only non-Hispanic white and Black voters from the matched sample for the interaction models. All models include all covariates along with jail and week-of-year fixed effects; robust standard errors are two-way clustered at the jail and week-of-year level.

Table \ref{tab:race_cov_0.75} reports estimates. The effect of jail incarceration on voting from jail in 2020 was significantly larger for Black registered voters, relative to white registered voters. Including all individual- and booking-level covariates along with jail and week fixed effects, white registered voters who were booked into county jails during 2020 voting days for the full set of voting days were on average 32.6\% -- 60.2\% less likely to have voted in 2020, relative to white registered voters who were booked into the same jails within 7 -- 42 days after Election Day (avg. effect magnitude = 39.8\%; p $<$ 0.01). The estimated negative effect of incarceration on voting from jail was 38.2\% - 55.0\% larger for Black registered voters (avg. effect magnitude = 45.9\%; p $<$ 0.05). The average total effect magnitude for Black registered voters of being booked into a county jail during 2020 voting days for the full set of voting days was a 77.9\% decrease in turnout, relative to Black registered voters booked into the same jails within 7 -- 42 days after Election Day.

The L2 race and ethnicity categories include both reported and predicted race/ethnicity. Table \ref{tab:race_cov_reporting_0.75} restricts the analyses reported in Table \ref{tab:race_cov_0.75} to only those states that report race/ethnicity to L2, excluding observations where race/ethnicity is predicted by L2 using other features available in the data. Effect magnitudes and significance levels are similar for both white and Black registered voters.

\subsection{Match Probability $p>0.95$}

The sample of linked records used in our main analyses have match probabilities that exceed 0.75. In Appendix \ref{app:a} we also replicate our primary analyses using only linked records with match probabilities that exceed 0.95. Figure \ref{fig:pvals_0.95} reports for this sample estimates of balance on individual- and booking-level characteristics for post-Election Day control windows ranging between 7 and 42 days and treatment windows containing legal voting days in 2020. Table \ref{tab:random_0.95} reports for this sample estimates of balance on individual- and booking-level characteristics for post-Election Day control windows ranging between 7 and 42 days and the largest treatment windows for which the joint F-test p-values for that treatment window and for all smaller treatment windows are greater than 0.10. Balanced samples are considerably smaller than those for linked records with match probabilities that exceed 0.75. Figure \ref{fig:early_0.95} reports the distribution of ballot return and the date of the start of the smallest treatment window among the samples in Table \ref{tab:random_0.95}. Table \ref{tab:stats_0.95} reports summary statistics for the balanced samples reported in Table \ref{tab:random_0.95}.

Table \ref{tab:turnout_0.95} reports estimates of the effect of jail incarceration on the voting behavior of registered voters, for the sample of linked records with match probabilities exceeding 0.95. Effect magnitudes are larger than those reported for the sample of linked records with match probabilities exceeding 0.75. In models that include all individual- and booking-level covariates along with jail and week fixed effects, registered voters in this sample who were booked into county jails during 2020 voting days in their state for any length of time were on average 3.9 -- 4.5 percentage points or 13.1\% -- 14.5\% less likely to have voted in 2020, relative to registered voters who were booked into the same jails within 7 -- 42 days after Election Day (avg. effect magnitude = 13.9\%; p $<$ 0.01).

In the balanced samples that have coverage on registered voters who were incarcerated during the full span of 2020 voting days in their states (excluding the balanced sample including only 7 days pre- and post-Election Day), effect magnitudes are again larger than those reported for the sample of linked records with match probabilities exceeding 0.75. Including all individual- and booking-level covariates along with jail and week fixed effects, registered voters in this sample who were booked into county jails during 2020 voting days in their state for the full duration of voting days were on average 21.3 - 30 percentage points or 67.9\% -- 98\% less likely to have voted in 2020, relative to registered voters who were booked into the same jails within 14 -- 42 days after Election Day (avg. effect magnitude = 80.8\%; p $<$ 0.01). 

Table \ref{tab:placebo_0.95} reports no consistent effects of pre-Election Day jail incarceration in 2020 on registered voters' turnout in 2016 or 2012, for the sample of linked records with match probabilities exceeding 0.95. The effects of 2020 jail incarceration on 2016 turnout are very slightly positive and significant in two out of six models at p $<$ 0.05; the effects of 2020 jail incarceration on 2012 turnout are consistently insignificant. The effects of 2020 jail incarceration on 2020 turnout among registered voters who were also registered to vote in 2016 and/or 2012 are consistently negative, large, and significant, with magnitudes consistent with those observed for the sample of linked records with match probabilities exceeding 0.75.

\subsection{The Effect of Jail Incarceration on Registration}

Jail incarceration may also affect voting behavior by reducing the ability of jailed individuals to register to vote. If this were the case in 2020, then our estimates of the effect of jail incarceration on the voting behavior of registered voters in 2020 understate that effect.

In Appendix \ref{app:b} we replicate our analyses for the full sample of jailed individuals during our time periods of interest, both registered and unregistered. Figure \ref{fig:pvals_0.95_full} reports for this sample estimates of balance on individual- and booking-level characteristics for post-Election Day control windows ranging between 7 and 42 days and treatment windows containing legal voting days in 2020. Individual- and booking-level characteristics for this sample are sourced only from the JDI database in order to have coverage on both unregistered and registered voters. Table \ref{tab:random_0.95_full} reports the balance tests for the largest treatment windows for which the joint F-test p-values for that treatment window and for all smaller treatment windows are greater than 0.10. These windows are considerably smaller than the windows within which we obtain balanced samples of registered voters, limiting our ability to make inferences about the impacts of being incarcerated for the full span of legal voting days in 2020.

Table \ref{tab:stats_0.95_full} reports summary statistics for the balanced samples reported in Table \ref{tab:random_0.95_full}. Pre-Election Day voter registration rates among individuals who were booked into county jails within 7 -- 42 days after 2020 Election Day ranged between 29\% and 29.4\% in our samples. 2020 turnout rates in these same post-Election day samples, unconditional on registration status, ranged between 8.1\% and 8.4\%. 

Table \ref{tab:match_in_0.95_full} reports estimates of the effect of jail incarceration during 2020 voting days on the probability that an individual was registered to vote by Election Day 2020 (whether a jailed individual matched to L2 records with match probability $p>0.95$ and was registered to vote before Election Day 2020), for the balanced samples in Table \ref{tab:random_0.95_full}. In models that include all individual- and booking-level covariates along with jail and week fixed effects, there is a very small positive effect (0.02\%) of being jailed for any length of time before Election Day on the probability of being registered to vote by Election Day ($p<0.05$). The sign of the effect flips when incarceration is measured as proportion of voting days incarcerated.  Estimates of the effects of jail incarceration on registration behavior in 2020 using this continuous measure of incarceration are large, negative, and significant at p $<$ 0.01. However, because no balanced samples that include both registered and unregistered voters have coverage on individuals who were incarcerated during the full span of 2020 voting days in their states, we refrain from interpreting the effect magnitudes for this measure of incarceration. 

Table \ref{tab:turnout_0.95_full} reports estimates of the effect of jail incarceration during 2020 voting days on turnout unconditional on registration status, for the balanced samples in Table \ref{tab:random_0.95_full}. Effect magnitudes for the impacts of being incarcerated for any length of time before Election Day are comparable to those estimated only on samples of registered voters. In models that include all individual- and booking-level covariates along with jail and week fixed effects, individuals in this sample who were booked into county jails during 2020 voting days in their state for any length of time were on average 0.6 -- 0.7 percentage points or 7.2\% -- 8.3\% less likely to have voted in 2020, relative to individuals who were booked into the same jails within 7 -- 42 days after Election Day (avg. effect magnitude = 7.8\%; p $<$ 0.01). Estimates of the effects of jail incarceration on voting behavior in 2020 using the continuous measure of incarceration are large, negative, and significant at p $<$ 0.01. Again, however, because no balanced samples that include both registered and unregistered voters have coverage on individuals who were incarcerated during the full span of 2020 voting days in their states, we refrain from interpreting the effect magnitudes for this measure of incarceration. 

\section{Conclusion}

In every state, otherwise voting-eligible individuals incarcerated pretrial or to serve a misdemeanor sentence remain legally entitled to vote while incarcerated \citep{SP_20, SP_July22}. The vast majority of the approximately 650,000 individuals incarcerated in county jails on any given day are being held pretrial or to serve misdemeanor sentences \citep{whole_pie}. Anecdotal evidence suggests, however, that many of those incarcerated in county jails are not being given adequate opportunities to exercise their right to vote \citep{SP_20}. Concerns have also been raised about possible racial disparities in ballot access for those incarcerated in county jails \citep{SP_July22}. Yet we lack reliable causal estimates of the impacts of being incarcerated in a county jail on the exercise of the right to vote, and of any racial heterogeneity in those impacts.

We leverage new data on daily individual-level jail records and exploit the timing of incarceration to estimate the causal effects of jail incarceration on voting from jail in 2020. We find that registered voters booked into county jails for the full duration of 2020 voting days were on average 46\% less likely to vote in 2020, relative to registered voters booked into the same jails within 7 -- 42 days after Election Day. The estimated negative effect of incarceration on voting from jail was much larger for Black registered voters, who were 78\% less likely to vote in 2020 if booked into county jails for the full duration of 2020 voting days, relative to Black registered voters booked into the same jails just after Election Day. Placebo tests indicate no effects of 2020 jail incarceration on the 2012 or 2016 turnout of registered voters. We find few effects of jail incarceration on voter registration in 2020, and effect sizes of comparable magnitude for turnout unconditional on registration status. Our findings reveal the pressing need to enable voting-eligible incarcerated individuals to exercise their constitutional right to vote, and to address troubling racial disparities in the effect of jail incarceration on the exercise of the right to vote. 

\end{onehalfspacing}

\newpage
\vfill
\clearpage
\bibliographystyle{apsr.bst}
\bibliography{sample.bib}

\newpage

\clearpage

\begin{center}
\section*{Tables and Figures}
\end{center}

\clearpage

\clearpage

\begin{center}
\captionof{table}{Balance Tests}\label{tab:random_0.75}
\begin{adjustwidth}{-1.6cm}{}
\begin{threeparttable}
\begin{tabular}{llllllll}
\toprule
 & \textbf{Control:} & \textbf{7 days} & \textbf{14 days} & \textbf{21 days} & \textbf{28 days} & \textbf{35 days} & \textbf{42 days}\\
 & \textbf{Treatment:} & \textbf{60 days} & \textbf{53 days} & \textbf{60 days} & \textbf{17 days} & \textbf{31 days} & \textbf{60 days}\\
\midrule
Age (L2) &  & -0.000 & -0.000 & -0.000 & -0.000 & 0.000 & 0.000 \\
 &  & (0.000) & (0.000) & (0.000) & (0.000) & (0.000) & (0.000) \\
Black (L2) &  & 0.008 & 0.005 & 0.010 & 0.004 & 0.004 & 0.002 \\
 &  & (0.006) & (0.006) & (0.007) & (0.008) & (0.007) & (0.007) \\
White (L2) &  & 0.000 & -0.001 & 0.001 & -0.010 & -0.003 & -0.001 \\
 &  & (0.005) & (0.005) & (0.005) & (0.006) & (0.006) & (0.005) \\
Male (L2) &  & 0.002 & 0.007** & 0.006 & 0.006 & 0.003 & 0.004 \\
 &  & (0.003) & (0.004) & (0.004) & (0.005) & (0.004) & (0.004) \\
Democrat (L2) &  & -0.007* & -0.002 & -0.002 & -0.003 & 0.002 & -0.001 \\
 &  & (0.004) & (0.004) & (0.004) & (0.005) & (0.004) & (0.004) \\
Republican (L2) &  & -0.004 & -0.005 & -0.006 & -0.005 & -0.005 & -0.005 \\
 &  & (0.004) & (0.005) & (0.005) & (0.006) & (0.005) & (0.005) \\
Number of Charges (JDI) &  & 0.000 & 0.000 & 0.000 & -0.001 & -0.000 & -0.000 \\
 &  & (0.000) & (0.000) & (0.000) & (0.001) & (0.001) & (0.001) \\
Top Charge: DUI (JDI) &  & 0.003 & -0.015 & -0.009 & -0.002 & -0.003 & -0.011 \\
 &  & (0.010) & (0.011) & (0.011) & (0.014) & (0.012) & (0.011) \\
Top Charge: Drug (JDI) &  & 0.009 & -0.002 & -0.002 & -0.008 & 0.000 & -0.007 \\
 &  & (0.010) & (0.010) & (0.010) & (0.014) & (0.012) & (0.010) \\
Top Charge: Property (JDI) &  & 0.005 & -0.008 & -0.005 & -0.008 & -0.003 & -0.009 \\
 &  & (0.009) & (0.010) & (0.010) & (0.013) & (0.012) & (0.010) \\
Top Charge: Public Order (JDI) &  & 0.007 & -0.007 & -0.002 & -0.008 & 0.004 & -0.002 \\
 &  & (0.009) & (0.010) & (0.010) & (0.013) & (0.011) & (0.010) \\
Top Charge: Violent (JDI) &  & 0.007 & -0.014 & -0.007 & -0.015 & -0.011 & -0.016* \\
 &  & (0.008) & (0.010) & (0.010) & (0.012) & (0.011) & (0.010) \\
\midrule
Observations &  & 63000 & 70854 & 79854 & 57821 & 81683 & 103091 \\
Joint F-Test p-value &  & 0.614 & 0.101 & 0.372 & 0.162 & 0.131 & 0.182 \\
\bottomrule
\end{tabular}
\begin{tablenotes}
\small
\item \textbf{Note:} This table reports estimates of balance on individual- and booking-level characteristics for post-Election Day control windows ranging between 7 and 42 days and treatment windows containing legal voting days in 2020, for the sample of matched records (registered voters booked into county jails during the specified time periods) with match probability $p>0.75$. As reported in Figure \ref{fig:pvals_0.75}, for each control group window we initially tested for balance using treatment windows containing the full set of voting days in each state in 2020, sequentially dropping the earliest voting day in the sample and repeating the balance test until we reached a treatment window of 7 days prior to Election Day. This table reports balance for the largest treatment windows for which the joint F-test p-values for that treatment window and for all smaller treatment windows were greater than 0.10. The control window contains bookings beginning on November 4 through November 10. The outcome is whether a booking occurred during a treatment window (1) versus a control window (0). All models include jail-level fixed effects; robust standard errors clustered on jails are reported in parentheses. ***p $<$ 0.01, **p $<$ 0.05, *p $<$ 0.10.
\end{tablenotes}
\end{threeparttable}
\end{adjustwidth}
\end{center}

\clearpage

\begin{landscape}
\begin{center}
\captionof{table}{Summary Statistics}\label{tab:stats_0.75}
\begin{threeparttable}
\begin{tabular}{p{0.275\linewidth}  p{0.04\linewidth} p{0.04\linewidth}  p{0.04\linewidth}  p{0.04\linewidth}  p{0.04\linewidth}  p{0.04\linewidth}  p{0.04\linewidth}  p{0.04\linewidth}  p{0.04\linewidth}  p{0.04\linewidth}  p{0.04\linewidth}  p{0.04\linewidth}}
\toprule
\textbf{Treatment/Control:} & T & C & T & C & T & C & T & C & T & C & T & C \\
\textbf{Days:} & 60d & 7d & 53d & 14d & 60d & 21d & 17d & 28d & 31d & 35d & 60d & 42d \\
\midrule
Age (L2) & 37.178 & 37.365 & 37.175 & 37.391 & 37.178 & 37.275 & 37.152 & 37.217 & 37.187 & 37.241 & 37.178 & 37.232 \\
Black (L2) & 0.271 & 0.246 & 0.270 & 0.249 & 0.271 & 0.247 & 0.255 & 0.247 & 0.267 & 0.248 & 0.271 & 0.253 \\
White (L2) & 0.601 & 0.604 & 0.602 & 0.603 & 0.601 & 0.603 & 0.597 & 0.605 & 0.593 & 0.607 & 0.601 & 0.601 \\
Male (L2) & 0.734 & 0.736 & 0.733 & 0.731 & 0.734 & 0.733 & 0.737 & 0.734 & 0.735 & 0.734 & 0.734 & 0.734 \\
Democrat (L2) & 0.417 & 0.419 & 0.418 & 0.411 & 0.417 & 0.410 & 0.408 & 0.408 & 0.415 & 0.409 & 0.417 & 0.410 \\
Republican (L2) & 0.200 & 0.203 & 0.199 & 0.204 & 0.200 & 0.204 & 0.192 & 0.203 & 0.194 & 0.202 & 0.200 & 0.201 \\
Non-Partisan or Other Party (L2) & 0.383 & 0.378 & 0.383 & 0.385 & 0.383 & 0.386 & 0.400 & 0.389 & 0.391 & 0.389 & 0.383 & 0.389 \\
Number of Charges (JDI) & 2.507 & 2.539 & 2.499 & 2.502 & 2.507 & 2.501 & 2.466 & 2.520 & 2.494 & 2.504 & 2.507 & 2.514 \\
Top Charge: Violent (JDI) & 0.275 & 0.279 & 0.274 & 0.288 & 0.275 & 0.285 & 0.284 & 0.289 & 0.278 & 0.289 & 0.275 & 0.292 \\
Top Charge: Property (JDI) & 0.194 & 0.200 & 0.195 & 0.198 & 0.194 & 0.199 & 0.202 & 0.201 & 0.199 & 0.199 & 0.194 & 0.198 \\
Top Charge: Drug (JDI) & 0.138 & 0.131 & 0.138 & 0.128 & 0.138 & 0.131 & 0.127 & 0.128 & 0.133 & 0.129 & 0.138 & 0.131 \\
Top Charge: Public Order (JDI) & 0.298 & 0.288 & 0.298 & 0.287 & 0.298 & 0.287 & 0.290 & 0.287 & 0.294 & 0.286 & 0.298 & 0.283 \\
Top Charge: DUI (JDI) & 0.062 & 0.067 & 0.063 & 0.067 & 0.062 & 0.066 & 0.065 & 0.063 & 0.063 & 0.065 & 0.062 & 0.065 \\
Top Charge: Criminal Traffic (JDI) & 0.033 & 0.035 & 0.033 & 0.032 & 0.033 & 0.032 & 0.033 & 0.032 & 0.034 & 0.033 & 0.033 & 0.032 \\
Length of Stay (days) (JDI) & 37.458 & 37.207 & 37.527 & 35.561 & 37.458 & 35.104 & 33.885 & 34.566 & 36.559 & 34.279 & 37.458 & 35.559 \\
\midrule
Confined During Voting Days & 1.000 & 0.000 & 1.000 & 0.000 & 1.000 & 0.000 & 1.000 & 0.000 & 1.000 & 0.000 & 1.000 & 0.000 \\
Proportion of Voting Days Confined & 0.215 & 0.000 & 0.215 & 0.000 & 0.215 & 0.000 & 0.143 & 0.000 & 0.192 & 0.000 & 0.215 & 0.000 \\
\midrule
Turnout & 0.323 & 0.349 & 0.322 & 0.352 & 0.323 & 0.354 & 0.316 & 0.351 & 0.321 & 0.356 & 0.323 & 0.357 \\
\midrule
Observations & 53995 & 9005 & 53507 & 17347 & 53995 & 25859 & 23947 & 33874 & 40047 & 41636 & 53995 & 49096 \\
\bottomrule
\end{tabular}
\begin{tablenotes}
\small
\item \textbf{Note:} This table reports mean values of individual- and booking-level characteristics for the balanced samples reported in Table \ref{tab:random_0.75}, which comprise matched records (registered voters booked into county jails during the specified time periods) with match probability $p>0.75$, by treatment and control groups.
\end{tablenotes}
\end{threeparttable}
\end{center}
\end{landscape}

\clearpage

\begin{center}
\captionof{table}{The Effect of Incarceration on Voting From Jail}\label{tab:turnout_0.75}
\begin{threeparttable}
\begin{tabular}{lllll}
\toprule
\textbf{} & \textbf{Control:} & \textbf{7 days} & \textbf{14 days} & \textbf{21 days}\\
\textbf{} & \textbf{Treatment:} & \textbf{60 days} & \textbf{53 days} & \textbf{60 days}\\
\midrule
Confined During Voting Days &  & -0.029*** & -0.031*** & -0.031*** \\
 &  & (0.002) & (0.002) & (0.002) \\
with covariates &  & -0.030*** & -0.031*** & -0.032*** \\
 &  & (0.002) & (0.002) & (0.002) \\
\midrule
Proportion of Voting Days Confined &  & -0.169*** & -0.166*** & -0.166*** \\
 &  & (0.021) & (0.021) & (0.020) \\
with covariates &  & -0.151*** & -0.143*** & -0.142*** \\
 &  & (0.018) & (0.019) & (0.018) \\
\midrule
Mean Proportion of Voting Days Confined &  & 0.215 & 0.215 & 0.215 \\
Max. Proportion of Voting Days Confined &  & 1.000 & 1.000 & 1.000 \\
Mean Control Turnout &  & 0.349 & 0.352 & 0.354 \\
Observations &  & 63000 & 70854 & 79854 \\

\toprule
\textbf{} & \textbf{Control:} & \textbf{28 days} & \textbf{35 days} & \textbf{42 days}\\
\textbf{} & \textbf{Treatment:} & \textbf{17 days} & \textbf{31 days} & \textbf{60 days}\\
\midrule
Confined During Voting Days &  & -0.030*** & -0.032*** & -0.032*** \\
 &  & (0.002) & (0.002) & (0.002) \\
with covariates &  & -0.032*** & -0.033*** & -0.033*** \\
 &  & (0.002) & (0.002) & (0.002) \\
\midrule
Proportion of Voting Days Confined &  & -0.274*** & -0.204*** & -0.162*** \\
 &  & (0.043) & (0.023) & (0.019) \\
with covariates &  & -0.229*** & -0.175*** & -0.137*** \\
 &  & (0.040) & (0.020) & (0.018) \\
\midrule
Mean Proportion of Voting Days Confined &  & 0.143 & 0.192 & 0.215 \\
Max. Proportion of Voting Days Confined &  & 0.947 & 1.000 & 1.000 \\
Mean Control Turnout &  & 0.351 & 0.356 & 0.357 \\
Observations &  & 57821 & 81683 & 103091 \\
\bottomrule
\end{tabular}
\begin{tablenotes}
\small
\item \textbf{Note:} This table reports OLS estimates of (1) the effect of being booked into a county jail during 2020 voting days for any length of time on a registered voter's probability of voting in the 2020 election, with and without covariates; and (2) the effect of the proportion of 2020 voting days during which a registered voter booked into a county jail during voting days was confined on their probability of voting in the 2020 election, with and without covariates, for samples of matched records (registered voters booked into county jails during the specified time periods) with match probability $p>0.75$.  Models are estimated on samples as described in the notes to Table \ref{tab:random_0.75}. All models include jail and week-of-year fixed effects; robust standard errors two-way clustered at the jail and week-of-year level are reported in parentheses. ***p $<$ 0.01, **p $<$ 0.05, *p $<$ 0.10.
\end{tablenotes}
\end{threeparttable}
\end{center}

\clearpage

\begin{center}
\captionof{table}{Placebo Tests}\label{tab:placebo_0.75}
\begin{adjustwidth}{-0.9cm}{}
\begin{threeparttable}
\begin{tabular}{llllll}
\toprule
& & \textbf{2016 Placebo} & & \textbf{2012 Placebo} & \\
\midrule
\textbf{Control/Treatment: (7 days, 60 days)} &  & 2016 & 2020 & 2012 & 2020 \\
\midrule
Confined During Voting Days &  & 0.005 & -0.020*** & -0.012 & -0.024*** \\
 &  & (0.007) & (0.007) & (0.008) & (0.009) \\
Mean Control Turnout &  & 0.442 & 0.399 & 0.554 & 0.450 \\
Observations &  & 40693 & 40693 & 28388 & 28388 \\
 \midrule
\textbf{Control/Treatment: (14 days, 53 days)} &  & 2016 & 2020 & 2012 & 2020 \\
\midrule
Confined During Voting Days &  & 0.004 & -0.025*** & -0.007 & -0.030*** \\
 &  & (0.005) & (0.006) & (0.006) & (0.007) \\
Mean Control Turnout &  & 0.442 & 0.404 & 0.544 & 0.456 \\
Observations &  & 45732 & 45732 & 31873 & 31873 \\
 \midrule
\textbf{Control/Treatment: (21 days, 60 days)} &  & 2016 & 2020 & 2012 & 2020 \\
\midrule
Confined During Voting Days &  & 0.004 & -0.027*** & -0.009 & -0.033*** \\
 &  & (0.005) & (0.005) & (0.006) & (0.006) \\
Mean Control Turnout &  & 0.441 & 0.406 & 0.544 & 0.459 \\
Observations &  & 51471 & 51471 & 35862 & 35862 \\
 \midrule
\textbf{Control/Treatment: (28 days, 17 days)} &  & 2016 & 2020 & 2012 & 2020 \\
\midrule
Confined During Voting Days &  & 0.002 & -0.034*** & -0.011* & -0.040*** \\
 &  & (0.005) & (0.005) & (0.007) & (0.007) \\
Mean Control Turnout &  & 0.443 & 0.404 & 0.544 & 0.457 \\
Observations &  & 37135 & 37135 & 25644 & 25644 \\
 \midrule
\textbf{Control/Treatment: (35 days, 31 days)} &  & 2016 & 2020 & 2012 & 2020 \\
\midrule
Confined During Voting Days &  & 0.003 & -0.032*** & -0.009* & -0.037*** \\
 &  & (0.004) & (0.004) & (0.005) & (0.005) \\
Mean Control Turnout &  & 0.445 & 0.409 & 0.546 & 0.462 \\
Observations &  & 52505 & 52505 & 36358 & 36358 \\
 \midrule
\textbf{Control/Treatment: (42 days, 60 days)} &  & 2016 & 2020 & 2012 & 2020 \\
\midrule
Confined During Voting Days &  & 0.001 & -0.030*** & -0.010** & -0.036*** \\
 &  & (0.004) & (0.004) & (0.005) & (0.005) \\
Mean Control Turnout &  & 0.445 & 0.410 & 0.546 & 0.463 \\
Observations &  & 66386 & 66386 & 46045 & 46045 \\
\bottomrule
\end{tabular}
\begin{tablenotes}
\small
\item \textbf{Note:} This table reports placebo tests of the effects of 2020 jail incarceration on 2016 and 2012 voter turnout. The 2016 placebo samples include registered voters in each of our balanced samples who were registered to vote on both Election Day 2020 and Election Day 2016 (November 8, 2016). The 2016 placebo tests report OLS estimates of the effect of a 2020 booking occurring in a pre-Election Day 2020 window on a registered voter's probability of voting in the 2016 election, for the registered voters in each of the balanced samples reported in Table \ref{tab:random_0.75}. The 2012 placebo samples include registered voters in each of our balanced samples who were registered to vote on both Election Day 2020 and Election Day 2012 (November 6, 2012). The 2012 placebo tests report OLS estimates of the effect of a 2020 booking occurring in a pre-Election Day 2020 window on a registered voter's probability of voting in the 2012 election, for the registered voters in each of the balanced samples reported in Table \ref{tab:random_0.75}. All models include jail fixed effects; robust standard errors clustered at the jail level are reported in parentheses. ***p $<$ 0.01, **p $<$ 0.05, *p $<$ 0.10.
\end{tablenotes}
\end{threeparttable}
\end{adjustwidth}
\end{center}

\clearpage

\begin{center}
\captionof{table}{Racial Disparities in the Effect of Incarceration on Voting From Jail (With Covariates)}\label{tab:race_cov_0.75}
\begin{adjustwidth}{-0.5cm}{}
\begin{threeparttable}
\begin{tabular}{lllll}
\toprule
\textbf{} & \textbf{Control:} & \textbf{7 days} & \textbf{14 days} & \textbf{21 days}\\
\textbf{} & \textbf{Treatment:} & \textbf{60 days} & \textbf{53 days} & \textbf{60 days}\\
\midrule
Confined During Voting Days &  & -0.039*** & -0.042*** & -0.042*** \\
 &  & (0.004) & (0.005) & (0.005) \\
Confined $\times$ Black &  & -0.003 & 0.003 & -0.001 \\
 &  & (0.006) & (0.007) & (0.008) \\
\midrule
Proportion of Voting Days Confined &  & -0.135*** & -0.131*** & -0.128*** \\
 &  & (0.017) & (0.017) & (0.017) \\
Proportion Confined $\times$ Black &  & -0.060** & -0.050** & -0.055** \\
 &  & (0.024) & (0.023) & (0.023) \\
\midrule
Mean Proportion of Voting Days Confined Black &  & 0.209 & 0.209 & 0.209 \\
Max. Proportion of Voting Days Confined Black &  & 1.000 & 1.000 & 1.000 \\
Mean Proportion of Voting Days Confined White &  & 0.218 & 0.218 & 0.218 \\
Max. Proportion of Voting Days Confined White &  & 1.000 & 1.000 & 1.000 \\
Mean Control Turnout Black &  & 0.278 & 0.279 & 0.285 \\
Mean Control Turnout White &  & 0.377 & 0.380 & 0.380 \\
Observations &  & 54741 & 61414 & 69084 \\

\toprule
\textbf{} & \textbf{Control:} & \textbf{28 days} & \textbf{35 days} & \textbf{42 days}\\
\textbf{} & \textbf{Treatment:} & \textbf{17 days} & \textbf{31 days} & \textbf{60 days}\\
\midrule
Confined During Voting Days &  & -0.041*** & -0.042*** & -0.042*** \\
 &  & (0.003) & (0.004) & (0.004) \\
Confined $\times$ Black &  & -0.005 & -0.006 & -0.003 \\
 &  & (0.004) & (0.007) & (0.006) \\
\midrule
Proportion of Voting Days Confined &  & -0.228*** & -0.160*** & -0.125*** \\
 &  & (0.036) & (0.022) & (0.016) \\
Proportion Confined $\times$ Black &  & -0.118*** & -0.088*** & -0.054** \\
 &  & (0.030) & (0.014) & (0.022) \\
\midrule
Mean Proportion of Voting Days Confined Black &  & 0.124 & 0.178 & 0.209 \\
Max. Proportion of Voting Days Confined Black &  & 0.947 & 1.000 & 1.000 \\
Mean Proportion of Voting Days Confined White &  & 0.148 & 0.196 & 0.218 \\
Max. Proportion of Voting Days Confined White &  & 0.947 & 1.000 & 1.000 \\
Mean Control Turnout Black &  & 0.284 & 0.291 & 0.292 \\
Mean Control Turnout White &  & 0.379 & 0.382 & 0.383 \\
Observations &  & 49257 & 70041 & 89010 \\
\bottomrule
\end{tabular}
\begin{tablenotes}
\small
\item \textbf{Note:} This table reports OLS estimates of (1) the effect of being booked into a county jail during 2020 voting days for any length of time on a registered voter's probability of voting in the 2020 election, interacted with an indicator for whether the individual is Black (0 = non-Hispanic white), with covariates; and (2) the effect of the proportion of 2020 voting days during which a registered voter booked into a county jail during voting days was confined on their probability of voting in the 2020 election, interacted with an indicator for whether the individual is Black, with covariates. Models are estimated on samples as described in the notes to Table \ref{tab:random_0.75}, including only Black and white individuals. All models include jail and week-of-year fixed effects; robust standard errors two-way clustered at the jail and week-of-year level are reported in parentheses. ***p $<$ 0.01, **p $<$ 0.05, *p $<$ 0.10.
\end{tablenotes}
\end{threeparttable}
\end{adjustwidth}
\end{center}

\begin{center}
\begin{adjustwidth}{-0.5cm}{}
\begin{threeparttable}
\captionsetup{justification=centering}
\caption{Racial Disparities in the Effect of Incarceration on Voting From Jail (With Covariates) \\States Reporting Race}\label{tab:race_cov_reporting_0.75}
\begin{tabular}{lllll}
\toprule
\textbf{} & \textbf{Control:} & \textbf{7 days} & \textbf{14 days} & \textbf{21 days}\\
\textbf{} & \textbf{Treatment:} & \textbf{60 days} & \textbf{53 days} & \textbf{60 days}\\
\midrule
Confined During Voting Days &  & -0.044*** & -0.047*** & -0.047*** \\
 &  & (0.008) & (0.008) & (0.008) \\
Confined $\times$ Black &  & 0.008 & 0.012 & 0.008 \\
 &  & (0.008) & (0.008) & (0.010) \\
\midrule
Proportion of Voting Days Confined &  & -0.131*** & -0.126*** & -0.126*** \\
 &  & (0.029) & (0.029) & (0.028) \\
Proportion Confined $\times$ Black &  & -0.084*** & -0.070*** & -0.073*** \\
 &  & (0.020) & (0.021) & (0.021) \\
\midrule
Mean Proportion of Voting Days Confined Black &  & 0.197 & 0.196 & 0.197 \\
Max. Proportion of Voting Days Confined Black &  & 1.000 & 1.000 & 1.000 \\
Mean Proportion of Voting Days Confined White &  & 0.210 & 0.210 & 0.210 \\
Max. Proportion of Voting Days Confined White &  & 1.000 & 1.000 & 1.000 \\
Mean Control Turnout Black &  & 0.293 & 0.291 & 0.298 \\
Mean Control Turnout White &  & 0.379 & 0.384 & 0.385 \\
Observations &  & 28382 & 31616 & 35843 \\

\toprule
\textbf{} & \textbf{Control:} & \textbf{28 days} & \textbf{35 days} & \textbf{42 days}\\
\textbf{} & \textbf{Treatment:} & \textbf{17 days} & \textbf{31 days} & \textbf{60 days}\\
\midrule
Confined During Voting Days &  & -0.049*** & -0.049*** & -0.045*** \\
 &  & (0.007) & (0.007) & (0.007) \\
Confined $\times$ Black &  & 0.008 & 0.005 & 0.003 \\
 &  & (0.010) & (0.008) & (0.009) \\
\midrule
Proportion of Voting Days Confined &  & -0.335*** & -0.192*** & -0.121*** \\
 &  & (0.072) & (0.033) & (0.027) \\
Proportion Confined $\times$ Black &  & -0.076 & -0.083*** & -0.072*** \\
 &  & (0.075) & (0.025) & (0.021) \\
\midrule
Mean Proportion of Voting Days Confined Black &  & 0.111 & 0.165 & 0.197 \\
Max. Proportion of Voting Days Confined Black &  & 0.581 & 0.968 & 1.000 \\
Mean Proportion of Voting Days Confined White &  & 0.116 & 0.175 & 0.210 \\
Max. Proportion of Voting Days Confined White &  & 0.581 & 1.000 & 1.000 \\
Mean Control Turnout Black &  & 0.295 & 0.305 & 0.308 \\
Mean Control Turnout White &  & 0.382 & 0.385 & 0.388 \\
Observations &  & 24709 & 35881 & 45868 \\
\bottomrule
\end{tabular}
\begin{tablenotes}
\small
\item \textbf{Note:} This table reports OLS estimates of (1) the effect of being booked into a county jail during 2020 voting days for any length of time on a registered voter's probability of voting in the 2020 election, interacted with an indicator for whether the individual is Black (0 = non-Hispanic white), with covariates; and (2) the effect of the proportion of 2020 voting days during which a registered voter booked into a county jail during voting days was confined on their probability of voting in the 2020 election, interacted with an indicator for whether the individual is Black, with covariates. Models are estimated on samples as described in the notes to Table \ref{tab:random_0.75}, including only Black and white individuals, and including only states that report voter race to L2. All models include jail and week-of-year fixed effects; robust standard errors two-way clustered at the jail and week-of-year level are reported in parentheses. ***p $<$ 0.01, **p $<$ 0.05, *p $<$ 0.10.
\end{tablenotes}
\end{threeparttable}
\end{adjustwidth}
\end{center}

\clearpage

\newpage
\appendix
\renewcommand\thefigure{\thesection.\arabic{figure}}
\renewcommand\thetable{\thesection.\arabic{table}}
\counterwithin{table}{section}
\counterwithin{figure}{section}

\appendix
\begin{center}
\section{Appendix: Registered Voters; Match Probability $p>0.95$}\label{app:a}
\end{center}

\begin{center}
\begin{figure}[h]
\captionof{figure}{Balance Tests (Joint F-test P-values)\\Match Probability $p>0.95$}\label{fig:pvals_0.95}
\centering
\includegraphics[width=\textwidth]{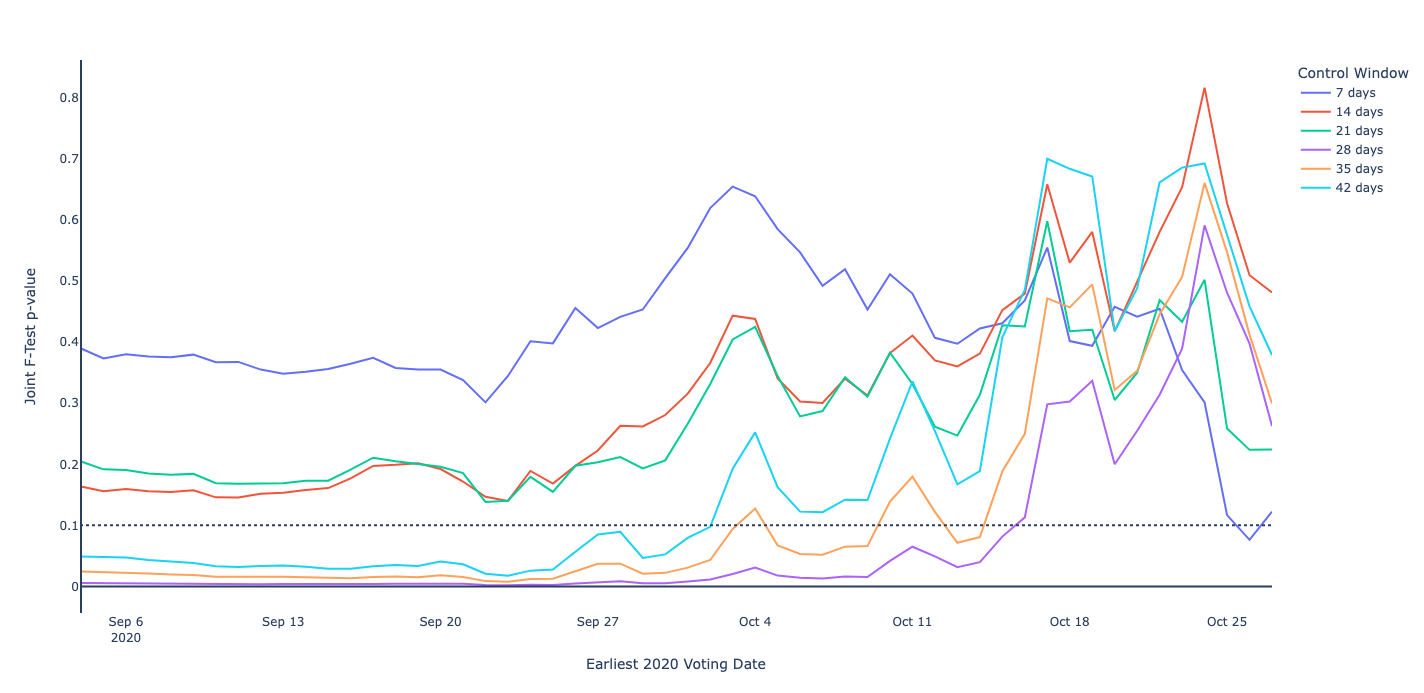}
\begin{tablenotes}
\small
\item \textbf{Note:} This figure reports estimates of balance on individual- and booking-level characteristics for post-Election Day control windows ranging between 7 and 42 days and treatment windows containing legal voting days in 2020, for the sample of matched records (registered voters incarcerated in county jails) with match probability $p>0.95$. For each post-Election Day control window we initially tested for balance using treatment windows containing the full set of voting days in each state in 2020, sequentially dropping the earliest voting day in the sample and repeating the balance test until we reached a treatment window of 7 days prior to Election Day.  
\end{tablenotes}
\end{figure}
\end{center}

\clearpage

\begin{center}
\begin{figure}[h]
\captionof{figure}{Distribution of Early Ballot Return\\Match Probability $p>0.95$}\label{fig:early_0.95}
\centering
\captionsetup{justification=centering}
\includegraphics[width=\textwidth]{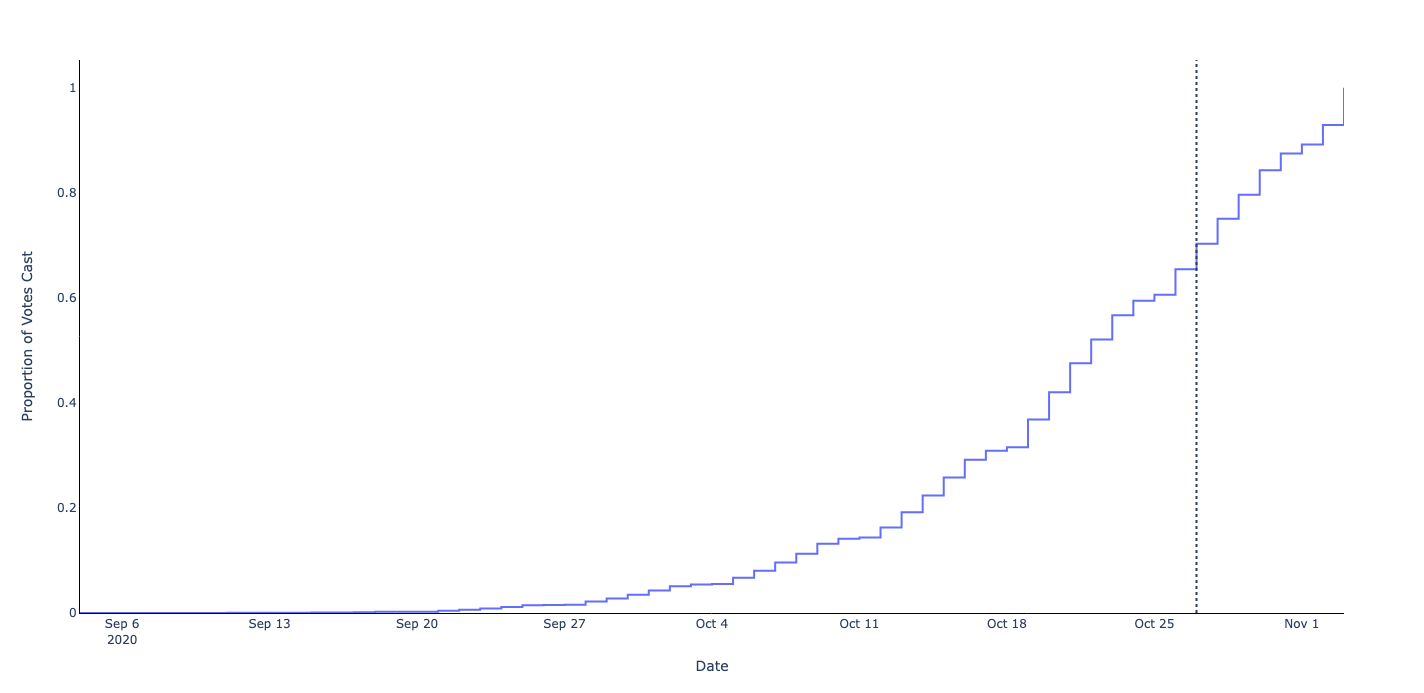}
\begin{tablenotes}
\small
\item \textbf{Note:} This figure reports the cumulative distribution of early ballot return in the L2 voter file data for 2020, where available. The smallest treatment window in the samples reported in Table \ref{tab:random_0.95} begins on October 27, indicated with a vertical dashed line. Approximately 30\% of 2020 ballots were returned after October 27.
\end{tablenotes}
\end{figure}
\end{center}

\clearpage
\begin{center}
\captionof{table}{Balance Tests\\Match Probability $p>0.95$}\label{tab:random_0.95}
\begin{adjustwidth}{-1.6cm}{}
\begin{threeparttable}
\begin{tabular}{llllllll}
\toprule
 & \textbf{Control:} & \textbf{7 days} & \textbf{14 days} & \textbf{21 days} & \textbf{28 days} & \textbf{35 days} & \textbf{42 days}\\
 & \textbf{Treatment:} & \textbf{7 days} & \textbf{60 days} & \textbf{60 days} & \textbf{18 days} & \textbf{19 days} & \textbf{31 days}\\
\midrule
Age (L2) &  & 0.000 & -0.000 & -0.000 & -0.000 & 0.000 & 0.000 \\
 &  & (0.000) & (0.000) & (0.000) & (0.000) & (0.000) & (0.000) \\
Black (L2) &  & 0.036** & 0.012* & 0.020** & 0.019* & 0.013 & 0.012 \\
 &  & (0.017) & (0.007) & (0.008) & (0.010) & (0.009) & (0.008) \\
White (L2) &  & 0.012 & 0.002 & 0.006 & -0.001 & -0.005 & 0.003 \\
 &  & (0.014) & (0.006) & (0.006) & (0.007) & (0.006) & (0.006) \\
 Male (L2) &  & -0.003 & 0.007* & 0.005 & 0.005 & 0.004 & 0.001 \\
 &  & (0.009) & (0.004) & (0.004) & (0.005) & (0.005) & (0.005) \\
Democrat (L2) &  & -0.026** & -0.005 & -0.007 & -0.008 & -0.010* & -0.005 \\
 &  & (0.011) & (0.005) & (0.005) & (0.006) & (0.005) & (0.005) \\
Republican (L2) &  & -0.000 & -0.005 & -0.005 & -0.005 & -0.005 & -0.008 \\
 &  & (0.012) & (0.006) & (0.006) & (0.007) & (0.007) & (0.006) \\
Number of Charges (JDI) &  & -0.001 & 0.000 & -0.000 & -0.001 & -0.001 & -0.000 \\
 &  & (0.001) & (0.000) & (0.000) & (0.001) & (0.001) & (0.001) \\
Top Charge: DUI (JDI) &  & 0.029 & -0.017 & -0.012 & -0.007 & 0.000 & -0.010 \\
 &  & (0.027) & (0.013) & (0.013) & (0.017) & (0.015) & (0.013) \\
Top Charge: Drug (JDI) &  & 0.052** & 0.000 & -0.001 & 0.002 & 0.010 & -0.002 \\
 &  & (0.026) & (0.012) & (0.012) & (0.015) & (0.015) & (0.013) \\
Top Charge: Property (JDI) &  & 0.044* & -0.004 & -0.003 & -0.002 & 0.008 & -0.005 \\
 &  & (0.026) & (0.011) & (0.012) & (0.015) & (0.014) & (0.013) \\
Top Charge: Public Order (JDI) &  & 0.041 & -0.006 & -0.003 & -0.004 & 0.009 & 0.000 \\
 &  & (0.025) & (0.011) & (0.011) & (0.014) & (0.014) & (0.012) \\
Top Charge: Violent (JDI) &  & 0.056** & -0.011 & -0.008 & -0.013 & -0.002 & -0.015 \\
 &  & (0.024) & (0.011) & (0.011) & (0.014) & (0.013) & (0.012) \\
\midrule
Observations &  & 15312 & 54252 & 60834 & 45375 & 52980 & 68785 \\
Joint F-Test p-value &  & 0.122 & 0.164 & 0.205 & 0.113 & 0.188 & 0.193 \\
\bottomrule
\end{tabular}
\begin{tablenotes}
\small
\item \textbf{Note:} This table reports estimates of balance on individual- and booking-level characteristics for post-Election Day control windows ranging between 7 and 42 days and treatment windows containing legal voting days in 2020,for the sample of matched records (registered voters incarcerated in county jails) with match probability $p>0.95$. As reported in Figure \ref{fig:pvals_0.95}, for each control group window we initially tested for balance using treatment windows containing the full set of voting days in each state in 2020, sequentially dropping the earliest voting day in the sample and repeating the balance test until we reached a treatment window of 7 days prior to Election Day. This table reports balance for the largest treatment windows for which the joint F-test p-values for that treatment window and for all smaller treatment windows were greater than 0.10. The control window contains bookings beginning on November 4 through November 10. The outcome is whether a booking occurred during a treatment window (1) versus a control window (0). All models include jail-level fixed effects; robust standard errors clustered on jails are reported in parentheses. ***p $<$ 0.01, **p $<$ 0.05, *p $<$ 0.10.
\end{tablenotes}
\end{threeparttable}
\end{adjustwidth}
\end{center}

\clearpage

\begin{landscape}
\begin{center}
\captionof{table}{Summary Statistics\\Match Probability $p>0.95$}\label{tab:stats_0.95}
\begin{threeparttable}
\begin{tabular}{p{0.275\linewidth}  p{0.04\linewidth} p{0.04\linewidth}  p{0.04\linewidth}  p{0.04\linewidth}  p{0.04\linewidth}  p{0.04\linewidth}  p{0.04\linewidth}  p{0.04\linewidth}  p{0.04\linewidth}  p{0.04\linewidth}  p{0.04\linewidth}  p{0.04\linewidth}}
\toprule
\textbf{Treatment/Control:} & T & C & T & C & T & C & T & C & T & C & T & C \\
\textbf{Days:} & 7d & 7d & 60d & 14d & 60d & 21d & 18d & 28d & 19d & 35d & 31d & 42d \\
\midrule
Age (L2) & 36.706 & 36.744 & 36.650 & 36.846 & 36.650 & 36.771 & 36.634 & 36.712 & 36.642 & 36.714 & 36.659 & 36.705 \\
Black (L2) & 0.269 & 0.258 & 0.289 & 0.260 & 0.289 & 0.260 & 0.271 & 0.261 & 0.272 & 0.261 & 0.284 & 0.267 \\
White (L2) & 0.595 & 0.593 & 0.587 & 0.594 & 0.587 & 0.592 & 0.586 & 0.593 & 0.586 & 0.596 & 0.581 & 0.589 \\
Male (L2) & 0.721 & 0.718 & 0.716 & 0.712 & 0.716 & 0.715 & 0.719 & 0.716 & 0.720 & 0.717 & 0.717 & 0.718 \\
Democrat (L2) & 0.404 & 0.423 & 0.422 & 0.413 & 0.422 & 0.414 & 0.409 & 0.412 & 0.409 & 0.413 & 0.419 & 0.414 \\
Republican (L2) & 0.177 & 0.184 & 0.182 & 0.188 & 0.182 & 0.187 & 0.176 & 0.186 & 0.176 & 0.185 & 0.177 & 0.185 \\
Non-Partisan or Other Party (L2) & 0.419 & 0.394 & 0.396 & 0.399 & 0.396 & 0.400 & 0.414 & 0.402 & 0.415 & 0.402 & 0.405 & 0.401 \\
Number of Charges (JDI) & 2.441 & 2.567 & 2.489 & 2.504 & 2.489 & 2.505 & 2.463 & 2.517 & 2.480 & 2.495 & 2.477 & 2.502 \\
Top Charge: Violent (JDI) & 0.298 & 0.283 & 0.284 & 0.296 & 0.284 & 0.295 & 0.292 & 0.301 & 0.291 & 0.299 & 0.287 & 0.302 \\
Top Charge: Property (JDI) & 0.193 & 0.195 & 0.190 & 0.191 & 0.190 & 0.193 & 0.197 & 0.194 & 0.197 & 0.194 & 0.194 & 0.192 \\
Top Charge: Drug (JDI) & 0.127 & 0.129 & 0.135 & 0.127 & 0.135 & 0.129 & 0.127 & 0.125 & 0.127 & 0.126 & 0.131 & 0.127 \\
Top Charge: Public Order (JDI) & 0.289 & 0.288 & 0.292 & 0.283 & 0.292 & 0.282 & 0.285 & 0.282 & 0.286 & 0.281 & 0.289 & 0.278 \\
Top Charge: DUI (JDI) & 0.064 & 0.068 & 0.065 & 0.071 & 0.065 & 0.069 & 0.067 & 0.067 & 0.067 & 0.068 & 0.065 & 0.068 \\
Top Charge: Criminal Traffic (JDI) & 0.029 & 0.036 & 0.033 & 0.032 & 0.033 & 0.032 & 0.032 & 0.032 & 0.032 & 0.033 & 0.033 & 0.032 \\
Length of Stay (days) (JDI) & 32.268 & 35.361 & 35.452 & 33.670 & 35.452 & 33.653 & 32.461 & 33.263 & 32.963 & 32.893 & 34.859 & 34.165 \\
\midrule
Confined During Voting Days & 1.000 & 0.000 & 1.000 & 0.000 & 1.000 & 0.000 & 1.000 & 0.000 & 1.000 & 0.000 & 1.000 & 0.000 \\
Proportion of Voting Days Confined & 0.088 & 0.000 & 0.206 & 0.000 & 0.206 & 0.000 & 0.146 & 0.000 & 0.153 & 0.000 & 0.186 & 0.000 \\
\midrule
Turnout & 0.264 & 0.297 & 0.269 & 0.305 & 0.269 & 0.308 & 0.267 & 0.306 & 0.267 & 0.308 & 0.269 & 0.311 \\
\midrule
Observations & 8060 & 7252 & 40962 & 13290 & 40962 & 19872 & 19414 & 25961 & 20568 & 32412 & 30515 & 38270 \\
\bottomrule
\end{tabular}
\begin{tablenotes}
\small
\item \textbf{Note:} This table reports mean values of individual- and booking-level characteristics for registered voters in the balanced samples reported in Table \ref{tab:random_0.95}, containing matched records (registered voters incarcerated in county jails) with match probability $p>0.95$, by treatment and control groups.
\end{tablenotes}
\end{threeparttable}
\end{center}
\end{landscape}

\clearpage

\begin{center}
\captionof{table}{The Effect of Incarceration on Voting From Jail\\Match Probability $p>0.95$}\label{tab:turnout_0.95}
\begin{threeparttable}
\begin{tabular}{lllll}
\toprule
\textbf{} & \textbf{Control:} & \textbf{7 days} & \textbf{14 days} & \textbf{21 days}\\
\textbf{} & \textbf{Treatment:} & \textbf{7 days} & \textbf{60 days} & \textbf{60 days}\\
\midrule
Confined During Voting Days &  & -0.036*** & -0.041*** & -0.042*** \\
 &  & (0.002) & (0.002) & (0.002) \\
with covariates &  & -0.039*** & -0.043*** & -0.043*** \\
 &  & (0.002) & (0.002) & (0.002) \\
\midrule
Proportion of Voting Days Confined &  & -0.710*** & -0.259*** & -0.256*** \\
 &  & (0.067) & (0.023) & (0.023) \\
with covariates &  & -0.592*** & -0.213*** & -0.209*** \\
 &  & (0.051) & (0.021) & (0.020) \\
\midrule
Mean Proportion of Voting Days Confined &  & 0.088 & 0.206 & 0.206 \\
Max. Proportion of Voting Days Confined &  & 0.421 & 1.000 & 1.000 \\
Mean Control Turnout &  & 0.297 & 0.305 & 0.308 \\
Observations &  & 15312 & 54252 & 60834 \\

\toprule
\textbf{} & \textbf{Control:} & \textbf{28 days} & \textbf{35 days} & \textbf{42 days}\\
\textbf{} & \textbf{Treatment:} & \textbf{18 days} & \textbf{19 days} & \textbf{31 days}\\
\midrule
Confined During Voting Days &  & -0.039*** & -0.039*** & -0.042*** \\
 &  & (0.002) & (0.002) & (0.002) \\
with covariates &  & -0.042*** & -0.043*** & -0.045*** \\
 &  & (0.002) & (0.002) & (0.002) \\
\midrule
Proportion of Voting Days Confined &  & -0.374*** & -0.349*** & -0.299*** \\
 &  & (0.045) & (0.045) & (0.025) \\
with covariates &  & -0.300*** & -0.278*** & -0.242*** \\
 &  & (0.037) & (0.037) & (0.021) \\
\midrule
Mean Proportion of Voting Days Confined &  & 0.146 & 0.153 & 0.186 \\
Max. Proportion of Voting Days Confined &  & 1.000 & 1.000 & 1.000 \\
Mean Control Turnout &  & 0.306 & 0.308 & 0.311 \\
Observations &  & 45375 & 52980 & 68785 \\
\bottomrule
\end{tabular}
\begin{tablenotes}
\small
\item \textbf{Note:} This table reports OLS estimates of (1) the effect of being booked into a county jail during 2020 voting days for any length of time on a registered voter's probability of voting in the 2020 election, with and without covariates; and (2) the effect of the proportion of 2020 voting days during which a registered voter booked into a county jail during voting days was confined on their probability of voting in the 2020 election, with and without covariates, for the sample of matched records (registered voters incarcerated in county jails) with match probability $p>0.95$. Models are estimated on samples as described in the notes to Table \ref{tab:random_0.95}. All models include jail and week-of-year fixed effects; robust standard errors two-way clustered at the jail and week-of-year level are reported in parentheses. ***p $<$ 0.01, **p $<$ 0.05, *p $<$ 0.10.
\end{tablenotes}
\end{threeparttable}
\end{center}

\clearpage

\begin{center}
\captionof{table}{Placebo Tests\\Match Probability $p>0.95$}\label{tab:placebo_0.95}
\begin{adjustwidth}{-0.9cm}{}
\begin{threeparttable}
\begin{tabular}{llllll}
\toprule
& & \textbf{2016 Placebo} & & \textbf{2012 Placebo} & \\
\midrule
\textbf{Control/Treatment: (7 days, 7 days)} &  & 2016 & 2020 & 2012 & 2020 \\
\midrule
Confined During Voting Days &  & 0.024** & -0.026** & 0.005 & -0.025** \\
 &  & (0.010) & (0.010) & (0.013) & (0.012) \\
Mean Control Turnout &  & 0.400 & 0.339 & 0.525 & 0.386 \\
Observations &  & 9575 & 9575 & 6488 & 6488 \\
 \midrule
\textbf{Control/Treatment: (14 days, 60 days)} &  & 2016 & 2020 & 2012 & 2020 \\
\midrule
Confined During Voting Days &  & 0.010* & -0.031*** & -0.002 & -0.034*** \\
 &  & (0.006) & (0.006) & (0.008) & (0.008) \\
Mean Control Turnout &  & 0.403 & 0.351 & 0.516 & 0.397 \\
Observations &  & 34029 & 34029 & 23132 & 23132 \\
 \midrule
\textbf{Control/Treatment: (21 days, 60 days)} &  & 2016 & 2020 & 2012 & 2020 \\
\midrule
Confined During Voting Days &  & 0.008 & -0.035*** & -0.003 & -0.041*** \\
 &  & (0.005) & (0.005) & (0.007) & (0.007) \\
Mean Control Turnout &  & 0.404 & 0.355 & 0.515 & 0.403 \\
Observations &  & 38117 & 38117 & 25914 & 25914 \\
 \midrule
\textbf{Control/Treatment: (28 days, 18 days)} &  & 2016 & 2020 & 2012 & 2020 \\
\midrule
Confined During Voting Days &  & 0.011** & -0.035*** & -0.002 & -0.039*** \\
 &  & (0.005) & (0.006) & (0.007) & (0.007) \\
Mean Control Turnout &  & 0.406 & 0.353 & 0.516 & 0.401 \\
Observations &  & 28325 & 28325 & 19123 & 19123 \\
 \midrule
\textbf{Control/Treatment: (35 days, 19 days)} &  & 2016 & 2020 & 2012 & 2020 \\
\midrule
Confined During Voting Days &  & 0.009* & -0.038*** & -0.002 & -0.042*** \\
 &  & (0.005) & (0.006) & (0.007) & (0.007) \\
Mean Control Turnout &  & 0.408 & 0.355 & 0.517 & 0.404 \\
Observations &  & 33079 & 33079 & 22325 & 22325 \\
 \midrule
\textbf{Control/Treatment: (42 days, 31 days)} &  & 2016 & 2020 & 2012 & 2020 \\
\midrule
Confined During Voting Days &  & 0.006 & -0.038*** & -0.003 & -0.042*** \\
 &  & (0.004) & (0.004) & (0.006) & (0.006) \\
Mean Control Turnout &  & 0.408 & 0.358 & 0.518 & 0.406 \\
Observations &  & 42934 & 42934 & 28987 & 28987 \\
\bottomrule
\end{tabular}
\begin{tablenotes}
\small
\item \textbf{Note:} This table reports placebo tests of the effects of 2020 jail incarceration on 2016 and 2012 voter turnout. The 2016 (2012) placebo samples include registered voters in each of our balanced samples who were registered to vote on both Election Day 2020 and Election Day 2016 (Election Day 2012). The 2016 (2012) placebo tests report OLS estimates of the effect of a 2020 booking occurring in a pre-Election Day 2020 window on a registered voter's probability of voting in the 2016 (2012) election, for the registered voters in each of the balanced samples reported in Table \ref{tab:random_0.95}. All models include jail fixed effects; robust standard errors clustered at the jail level are reported in parentheses. ***p $<$ 0.01, **p $<$ 0.05 , *p $<$ 0.10.
\end{tablenotes}
\end{threeparttable}
\end{adjustwidth}
\end{center}

\clearpage

\begin{center}
\section{Appendix: Registered and Unregistered Voters}\label{app:b}
\end{center}

\begin{center}
\begin{figure}[h]
\captionof{figure}{Balance Tests (Joint F-test P-values)\\Registered and Unregistered Voters}\label{fig:pvals_0.95_full}
\centering
\includegraphics[width=\textwidth]{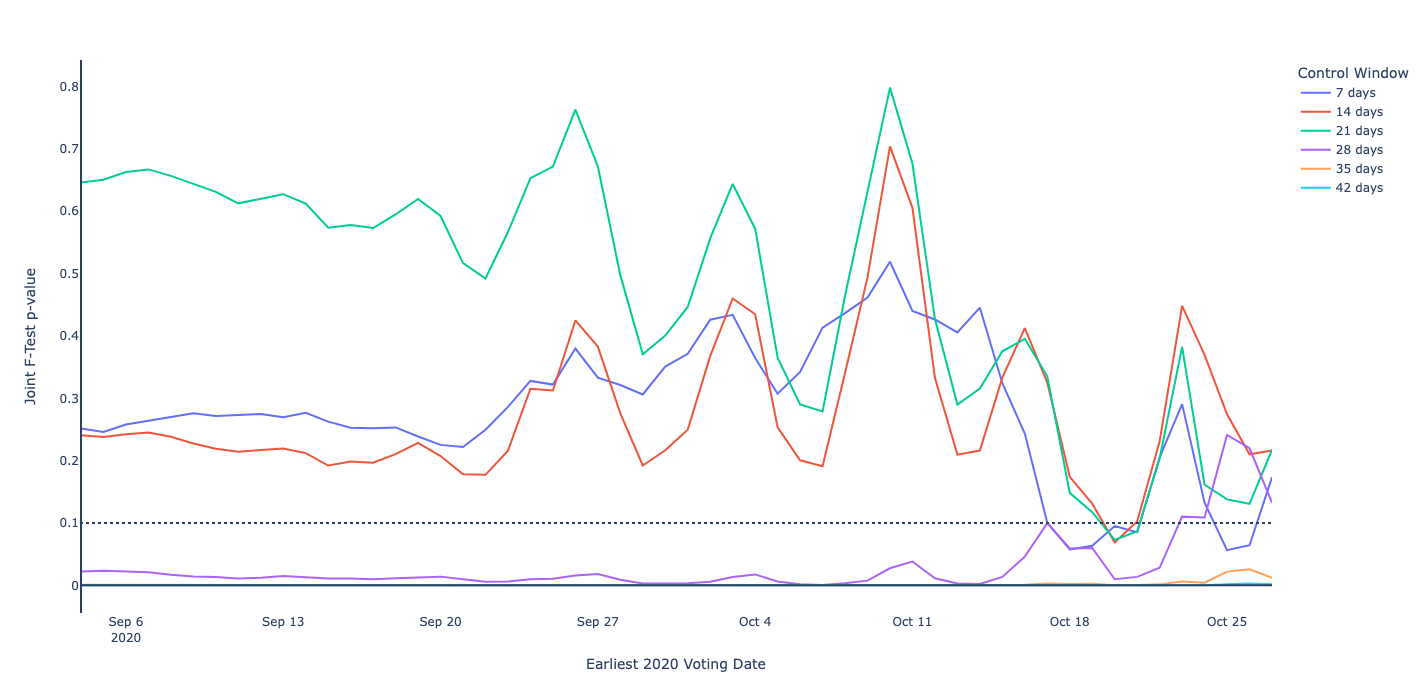}
\begin{tablenotes}
\small
\item \textbf{Note:} This figure reports estimates of balance on individual- and booking-level characteristics for post-Election Day control windows ranging between 7 and 42 days and treatment windows containing legal voting days in 2020, for the full sample of individuals booked into jail during this time period. For each post-Election Day control window we initially tested for balance using treatment windows containing the full set of voting days in each state in 2020, sequentially dropping the earliest voting day in the sample and repeating the balance test until we reached a treatment window of 7 days prior to Election Day.  
\end{tablenotes}
\end{figure}
\end{center}

\clearpage

\begin{center}
\captionof{table}{Balance Tests\\Registered and Unregistered Voters}\label{tab:random_0.95_full}
\begin{threeparttable}
\begin{tabular}{llllll}
\toprule
 & \textbf{Control:} & \textbf{7 days} & \textbf{14 days} & \textbf{21 days} & \textbf{28 days}\\
 & \textbf{Treatment:} & \textbf{7 days} & \textbf{13 days} & \textbf{12 days} & \textbf{11 days}\\
\midrule
Age (JDI) &  & 0.000 & 0.000 & 0.000 & 0.000 \\
 &  & (0.000) & (0.000) & (0.000) & (0.000) \\
Black (JDI) &  & 0.006 & 0.014 & 0.010 & 0.007 \\
 &  & (0.012) & (0.009) & (0.008) & (0.007) \\
White (JDI) &  & -0.005 & 0.005 & 0.004 & 0.003 \\
 &  & (0.011) & (0.009) & (0.008) & (0.007) \\
Male (JDI) &  & -0.001 & -0.001 & -0.000 & -0.000 \\
 &  & (0.006) & (0.005) & (0.004) & (0.004) \\
Number of Charges (JDI) &  & -0.001 & -0.000 & -0.000 & -0.001* \\
 &  & (0.000) & (0.001) & (0.001) & (0.000) \\
Top Charge: DUI (JDI) &  & 0.022 & -0.004 & -0.004 & -0.006 \\
 &  & (0.020) & (0.016) & (0.014) & (0.013) \\
Top Charge: Drug (JDI) &  & 0.032* & 0.021 & 0.012 & 0.003 \\
 &  & (0.020) & (0.015) & (0.013) & (0.012) \\
Top Charge: Property (JDI) &  & 0.033* & 0.014 & 0.005 & -0.005 \\
 &  & (0.018) & (0.014) & (0.013) & (0.012) \\
Top Charge: Public Order (JDI) &  & 0.029* & 0.018 & 0.012 & 0.003 \\
 &  & (0.018) & (0.014) & (0.012) & (0.011) \\
Top Charge: Violent (JDI) &  & 0.041** & 0.018 & 0.013 & -0.003 \\
 &  & (0.018) & (0.014) & (0.013) & (0.011) \\
\midrule
Observations &  & 38623 & 72096 & 86248 & 96446 \\
Joint F-Test p-value &  & 0.172 & 0.103 & 0.203 & 0.115 \\
\bottomrule
\end{tabular}
\begin{tablenotes}
\small
\item \textbf{Note:} This table reports estimates of balance on individual- and booking-level characteristics for post-Election Day control windows ranging between 7 and 28 days and treatment windows containing legal voting days in 2020, for the full sample of individuals booked into jail during this time period. As reported in Figure \ref{fig:pvals_0.95_full}, for each control group window we initially tested for balance using treatment windows containing the full set of voting days in each state in 2020, sequentially dropping the earliest voting day in the sample and repeating the balance test until we reached a treatment window of 7 days prior to Election Day. This table reports balance for the largest treatment windows for which the joint F-test p-values for that treatment window and for all smaller treatment windows were greater than 0.10.  The outcome is whether a booking occurred during a treatment window (1) versus a control window (0). All models include jail-level fixed effects; robust standard errors clustered on jails are reported in parentheses. ***p $<$ 0.01, **p $<$ 0.05, *p $<$ 0.10.
\end{tablenotes}
\end{threeparttable}
\end{center}

\clearpage

\begin{landscape}
\begin{center}
\captionof{table}{Summary Statistics\\Registered and Unregistered Voters}\label{tab:stats_0.95_full}
\begin{threeparttable}
\begin{tabular}{p{0.275\linewidth}  p{0.04\linewidth} p{0.04\linewidth}  p{0.04\linewidth}  p{0.04\linewidth}  p{0.04\linewidth}  p{0.04\linewidth}  p{0.04\linewidth}  p{0.04\linewidth}}
\toprule
\textbf{Treatment/Control:} & T & C & T & C & T & C & T & C \\
\textbf{Days:} & 7d & 7d & 13d & 14d & 12d & 21d & 11d & 28d \\
\midrule
Age (JDI) & 35.708 & 35.742 & 35.749 & 35.750 & 35.755 & 35.697 & 35.708 & 35.669 \\
Black (JDI) & 0.343 & 0.325 & 0.344 & 0.328 & 0.345 & 0.328 & 0.340 & 0.328 \\
White (JDI) & 0.582 & 0.594 & 0.582 & 0.592 & 0.581 & 0.592 & 0.584 & 0.592 \\
Male (JDI) & 0.776 & 0.776 & 0.772 & 0.773 & 0.773 & 0.773 & 0.773 & 0.774 \\
Number of Charges (JDI) & 2.716 & 2.778 & 2.739 & 2.777 & 2.744 & 2.777 & 2.723 & 2.812 \\
Top Charge: Violent (JDI) & 0.261 & 0.253 & 0.260 & 0.259 & 0.259 & 0.257 & 0.259 & 0.264 \\
Top Charge: Property (JDI) & 0.207 & 0.207 & 0.204 & 0.206 & 0.203 & 0.208 & 0.205 & 0.209 \\
Top Charge: Drug (JDI) & 0.142 & 0.145 & 0.145 & 0.144 & 0.145 & 0.145 & 0.143 & 0.141 \\
Top Charge: Public Order (JDI) & 0.321 & 0.318 & 0.322 & 0.315 & 0.323 & 0.315 & 0.320 & 0.313 \\
Top Charge: DUI (JDI) & 0.042 & 0.045 & 0.042 & 0.046 & 0.042 & 0.046 & 0.044 & 0.045 \\
Top Charge: Criminal Traffic (JDI) & 0.027 & 0.032 & 0.028 & 0.029 & 0.028 & 0.029 & 0.029 & 0.028 \\
Length of Stay (days) (JDI) & 41.127 & 44.463 & 43.036 & 42.640 & 42.681 & 42.187 & 40.931 & 41.502 \\
\midrule
Confined During Voting Days & 1.000 & 0.000 & 1.000 & 0.000 & 1.000 & 0.000 & 1.000 & 0.000 \\
Proportion of Voting Days Confined & 0.091 & 0.000 & 0.133 & 0.000 & 0.127 & 0.000 & 0.117 & 0.000 \\
\midrule
Registered by Election Day & 0.292 & 0.290 & 0.291 & 0.292 & 0.290 & 0.294 & 0.292 & 0.292 \\
Voted & 0.073 & 0.081 & 0.071 & 0.083 & 0.072 & 0.084 & 0.073 & 0.084 \\
\midrule
Observations & 20501 & 18122 & 37287 & 34809 & 34371 & 51877 & 30669 & 65777 \\
\bottomrule
\end{tabular}
\begin{tablenotes}
\small
\item \textbf{Note:} This table reports mean values of individual- and booking-level characteristics for individuals in the balanced samples reported in Table \ref{tab:random_0.95_full}, comprising both registered and unregistered voters booked into county jails during the specified time periods, by treatment and control groups.
\end{tablenotes}
\end{threeparttable}
\end{center}
\end{landscape}

\clearpage

\begin{center}
\captionof{table}{The Effect of Incarceration on Registering From Jail}\label{tab:match_in_0.95_full}
\begin{adjustwidth}{-0.8cm}{}
\begin{threeparttable}
\begin{tabular}{llllll}
\toprule
\textbf{} & \textbf{Control:} & \textbf{7 days} & \textbf{14 days} & \textbf{21 days} & \textbf{28 days}\\
\textbf{} & \textbf{Treatment:} & \textbf{7 days} & \textbf{13 days} & \textbf{12 days} & \textbf{11 days}\\
\midrule
Confined During Voting Days &  & 0.011*** & 0.011*** & 0.010*** & 0.011*** \\
 &  & (0.002) & (0.001) & (0.001) & (0.001) \\
with covariates &  & 0.007*** & 0.007*** & 0.006*** & 0.006*** \\
 &  & (0.002) & (0.001) & (0.001) & (0.001) \\
\midrule
Proportion of Voting Days Confined &  & -0.335*** & -0.278*** & -0.287*** & -0.296*** \\
 &  & (0.083) & (0.043) & (0.042) & (0.044) \\
with covariates &  & -0.184*** & -0.163*** & -0.168*** & -0.174*** \\
 &  & (0.068) & (0.032) & (0.030) & (0.032) \\
\midrule
Mean Proportion of Voting Days Confined &  & 0.091 & 0.133 & 0.127 & 0.117 \\
Max. Proportion of Voting Days Confined &  & 0.421 & 0.737 & 0.684 & 0.632 \\
Mean Control Registration Rate &  & 0.290 & 0.292 & 0.294 & 0.292 \\
Observations &  & 38623 & 72096 & 86248 & 96446 \\
\bottomrule
\end{tabular}
\begin{tablenotes}
\small
\item \textbf{Note:} This table reports OLS estimates of (1) the effect of being booked into a county jail during 2020 voting days for any length of time on an individual's probability of being registered to vote, with and without covariates; and (2) the effect of the proportion of 2020 voting days during which an individual booked into a county jail during voting days was confined on their probability of being registered to vote, with and without covariates, where registration is defined as matching to L2 records with match probability $p>0.95$. Models are estimated on samples as described in the notes to Table \ref{tab:random_0.95_full}. All models include jail and week-of-year fixed effects; robust standard errors two-way clustered at the jail and week-of-year level are reported in parentheses. ***p $<$ 0.01, **p $<$ 0.05, *p $<$ 0.10.
\end{tablenotes}
\end{threeparttable}
\end{adjustwidth}
\end{center}

\clearpage

\begin{center}
\captionsetup{justification=centering}
\captionof{table}{The Effect of Incarceration on Voting From Jail\\Unconditional on Registration Status}\label{tab:turnout_0.95_full}
\begin{adjustwidth}{-0.8cm}{}
\begin{threeparttable}
\begin{tabular}{llllll}
\toprule
\textbf{} & \textbf{Control:} & \textbf{7 days} & \textbf{14 days} & \textbf{21 days} & \textbf{28 days}\\
\textbf{} & \textbf{Treatment:} & \textbf{7 days} & \textbf{13 days} & \textbf{12 days} & \textbf{11 days}\\
\midrule
Confined During Voting Days &  & -0.005*** & -0.005*** & -0.005*** & -0.005*** \\
 &  & (0.001) & (0.001) & (0.001) & (0.001) \\
with covariates &  & -0.006*** & -0.007*** & -0.007*** & -0.007*** \\
 &  & (0.001) & (0.001) & (0.001) & (0.001) \\
\midrule
Proportion of Voting Days Confined &  & -0.278*** & -0.206*** & -0.217*** & -0.239*** \\
 &  & (0.040) & (0.033) & (0.034) & (0.031) \\
with covariates &  & -0.215*** & -0.154*** & -0.162*** & -0.181*** \\
 &  & (0.033) & (0.028) & (0.028) & (0.025) \\
\midrule
Mean Proportion of Voting Days Confined &  & 0.091 & 0.133 & 0.127 & 0.117 \\
Max. Proportion of Voting Days Confined &  & 0.421 & 0.737 & 0.684 & 0.632 \\
Mean Control Turnout &  & 0.081 & 0.083 & 0.084 & 0.084 \\
Observations &  & 38623 & 72096 & 86248 & 96446 \\
\bottomrule
\end{tabular}
\begin{tablenotes}
\small
\item \textbf{Note:} This table reports OLS estimates of (1) the effect of being booked into a county jail during 2020 voting days for any length of time on an individual's probability of voting in the 2020 election, with and without covariates; and (2) the effect of the proportion of 2020 voting days during which an individual booked into a county jail during voting days was incarcerated on their probability of voting in the 2020 election, with and without covariates, unconditional on registration status. Models are estimated on samples as described in the notes to Table \ref{tab:random_0.95_full}. All models include jail and week-of-year fixed effects; robust standard errors two-way clustered at the jail and week-of-year level are reported in parentheses. ***p $<$ 0.01, **p $<$ 0.05, *p $<$ 0.10.
\end{tablenotes}
\end{threeparttable}
\end{adjustwidth}
\end{center}

\clearpage

\newpage

\begin{center}
\section{Appendix: Data and Matching}\label{app:c}
\end{center}

\begin{onehalfspacing}

\subsection{Jail Data Initiative Booking Records}

Our data on jail bookings are sourced from the Jail Data Initiative (JDI) NoSQL database of booking records\footnote{For detailed documentation of the JDI pipeline and database architecture, see \url{https://jaildatainitiative.org/documentation}.} We initially sampled all individuals in the JDI database with jail booking dates from August 5, 2020 to February 1, 2021. These dates span 90 days on each side of 2020 Election Day (November 3, 2020) and include all legal voting days in 2020 (the earliest of which was September 4, 2020 in North Carolina, for absentee mail-in voting, 60 days prior to Election Day), and a post-Election Day period encompassing all control windows we employ. This sample includes 944,985 bookings. For each of these bookings we retrieved individual- and booking-level standardized fields from the JDI database as detailed below.

\textbf{\textit{Name}} \space Jail rosters may report a single name string, and/or first name, middle name, middle initial, and/or surname. In order to compare name components as part of the match process, we split names into components using the TupiLabs Java HumanNameParser package\footnote{\url{https://github.com/tupilabs/HumanNameParser.java}.}. Additionally, we create soundex codes for surnames, using the Apache Soundex Java class\footnote{\url{https://commons.apache.org/proper/commons-codec/apidocs/org/apache/commons/codec/language/Soundex.html}.}.

\textbf{\textit{Length of Stay}} \space Length of stay is used as a booking-level covariate in analyses of the effects of jail incarceration. Among the initial sample of 944,985 bookings, length of stay ranges from 1 to 736 days, and the mean length of stay is 44.48 days. Length of stay is used as a booking-level covariate in analyses of the effects of jail incarceration.

\textbf{\textit{Age}} \space Jail rosters may report age, year of birth, and/or date of birth. The JDI booking data report a standardized age field constructed from these original fields. Among the initial sample of 944,985 bookings, age ranges from 13 to 95, and the mean age is 35.7. The proportion without reported age is 0.19. Age is used as an individual-level covariate in analyses that include both unregistered and registered voters. We exclude from those samples booking records without reported age.

\textbf{\textit{Gender}} \space Jail rosters may report sex and/or gender. The JDI database reports a standardized sex-gender field from these original fields. Values (e.g., ``F", ``Female", ``FEM", etc.) are manually encoded as male, female, trans, non-binary or unknown. Among the initial sample of 944,985 bookings, the proportion without reported or encoded sex-gender is 0.24. Among those with encoded sex-gender, the proportion male is 0.78 and the proportion female is 0.23. 17 records (0.00) are encoded as non-binary or trans; these are re-categorized as unknown for matching with L2 voter records. Encoded sex-gender is used as an individual-level covariate in analyses that include both unregistered and registered voters. We exclude from those samples booking records without encoded sex-gender.

\textbf{\textit{Number of Charges}} \space Jail rosters may or may not report charges for each booking; charges may appear as a single string in an unparseable format. Among the initial sample of 944,985 bookings, the proportion with no determinable number of charges is 0.10. Among those with parseable charges, the mean number of charges per booking is 2.6. Number of charges is used as a booking-level covariate in analyses of the effects of jail incarceration. We exclude from all samples booking records without parseable charge records.

\textbf{\textit{Top Charge}} \space For each booking with parseable charge description(s) as specified above, the JDI database reports the highest-probability-match Uniform Crime Classification Standard (UCCS) code from the Text-Based Offense Classification (TOC) algorithm developed by the Criminal Justice Administrative Records System at the University of Michigan\footnote{For detailed information about the CJARS TOC model, see \url{https://cjars-toc.isr.umich.edu/}.}. In some cases, the charge string is not sufficiently legible for the model to produce a match classification. The JDI database assigns a ``top" charge type for the booking from among the least granular classification codes (``offense type" codes, with values 1-6 corresponding to the charge categories outlined below), using a hierarchical ordering of classification codes by severity. This order of severity, from most to least serious, is: violent, property, drug, public order, DUI, criminal traffic. Among the initial sample of 944,985 bookings, the proportion with no determinable top charge type is 0.2. Among those with encoded top charge types, the proportions of each type are: violent (0.21), property (0.17), drug (0.12), public order (0.25), DUI (0.04), and criminal traffic (0.03). Top charge indicators are used as booking-level covariates in analyses of the effects of jail incarceration. We exclude from all samples booking records without parseable charge records.

\textbf{\textit{County FIPS Code}} \space The JDI database reports the unique five-digit Federal Information Processing Standard Publication 6-4 (FIPS) county-identifying code from the Bureau of Justice Statistics 2013 Census of Jails corresponding to the county in which a jail is located.\footnote{\url{https://nvlpubs.nist.gov/nistpubs/Legacy/FIPS/fipspub6-4.pdf}, \url{https://doi.org/10.3886/ICPSR36128.v4}.} Among the initial sample of 944,985 bookings, there are 985 unique county FIPS codes.

\subsection{L2 Voter Records}

We sourced voter records from the L2 voter files licensed to New York University. The records used here were pulled in Spring 2021. The voter records from which we derive match candidates include 195,655,326 registered voters in the 42 states for which we have jail booking records.

\textbf{\textit{Age}} \space Among the initial sample of 195,655,326 voters, age ranges from 18 to 100, and the mean age is 50.07. The proportion without reported age is 0.02. L2 age is used as an individual-level covariate in analyses of the effects of jail incarceration on the voting behavior of registered voters. We exclude from those samples records without L2 age.

\textbf{\textit{Gender}} \space Among the initial sample of 195,655,326 voters, the proportion without reported gender is 0.01. Among those with reported gender, the proportion male is 0.47 and the proportion female is 0.53. L2 gender is used as an individual-level covariate in analyses of the effects of jail incarceration on the voting behavior of registered voters. We exclude from those samples records without L2 gender.

\textbf{\textit{Race}} \space Among the initial sample of 195,655,326 voters, the broadest L2 race/ethnicity field may take one of the following values: ``East and South Asian," ``European," ``Hispanic and Portuguese," ``Likely African-American," or ``Other." In states where these fields are not reported, L2 uses an algorithmic process to predict race/ethnicity, relying on a proprietary database of name-ethnicity mappings, with additional census block and secondary in-block surname assessment. We re-categorize L2 race/ethnicity into one of three possible values: (non-Hispanic) ``white" if L2 ethnicity value is ``European," (non-Hispanic) ``Black" if L2 ethnicity value is ``Likely African-American," and ``Other" if L2 ethnicity value is ``East and South Asian," ``Hispanic and Portuguese" or ``Other." Among our sample of voters, the proportion with unknown/unpredicted race is 0.09. Among those with known/predicted race, the proportion white is 0.68, the proportion Black is 0.12, and the proportion Other is 0.20. L2 race/ethnicity is used as an individual-level covariate in analyses of the effects of jail incarceration on the voting behavior of registered voters. We exclude from those samples records without L2 race/ethnicity.

\textbf{\textit{Party}} \space L2 reports voter party registration data where available. Where party registration data are not available, L2 models ``likely'' party identification (\url{https://www.l2datamapping.com/help?ds=VM_US#data__party}). The party registration/identification field may take one of 57 values or ``unknown.'' We re-categorize the party registration/identification field into one of three possible values: ``Democratic," ``Republican", and ``Non-Partisan or Other". Among our sample of voters, the proportion with unknown party registration/identification is 0.00. Among those with known/predicted party, the proportion Democratic is 0.41, the proportion Republican is 0.32, and the proportion Non-Partisan or Other is 0.28. L2 party registration/identification is used as an individual-level covariate in analyses of the effects of jail incarceration on the voting behavior of registered voters. We exclude from those samples records without L2 party registration/identification.

\textbf{\textit{Registration Date}} \space Voter registration date is used for sample inclusion in turnout analysis. L2 reports a calculated registration date field that reduces missingness in reported registration date, and captures the earliest registration date for an individual, since official registration dates are often reset when individuals move between voting districts. Among the initial sample of 195,655,326 voters, the mean calculated voter registration date is March 6, 1990.

\textbf{\textit{Voting Indicators}} \space L2 reports binary fields that indicate whether registered voters voted. We record such fields for the 2020, 2016 and 2012 general elections. Among the initial sample of 195,464,995 registered voters, the mean registered voter turnout for the 2020 general election was 0.75, the mean registered voter turnout for the 2016 general election was 0.59, and the mean registered voter turnout for the 2012 general election was 0.50.

\textbf{\textit{County FIPS Code}} \space Among the initial sample of 195,655,326 voters there are 3,004 unique county FIPS codes.

\subsection{Probabilistic Record Linkage}

Using the pools of booking and voter records described above, we conduct probabilistic record linkage using the Fellegi-Sunter/Expectation Maximization linkage model enumerated by \cite{mcdonough} and \cite{enamorado}.

Match-candidate booking rows appear as follows (note: full name and soundex code are used for blocking and initialization, but not for matching):

\begin{center}
\begin{tabular}{ |c||c|c|c|c|c|c|:c:c:| } 
 \hline
 Booking ID & FIPS & Age & Gender & First & Middle & Last & Full Name & Soundex \\ 
 \hline
 12a3b... & 11217 & 30 & & Jane & A & Doe & Jane A Doe & D000 \\
 \hline
\end{tabular}
\end{center}

Match-candidate voter rows appear as follows:

\begin{center}
\begin{tabular}{ |c||c|c|c|c|c|c|:c:c:| } 
 \hline
 Voter ID & FIPS & Age & Gender & First & Middle & Last & Full Name & Soundex \\ 
 \hline
 LALAL1... & 11217 & 29 & M & John & Adam & Doe & John Adam Doe & D000 \\
 \hline
\end{tabular}
\end{center}

We create blocks for matching as below. Importantly, we only run our match program within-state; if an individual is booked in Alabama but is registered in the Wyoming voter file, we do not consider them for matching. 

\begin{enumerate}
  \item For each unique age value among match-candidate bookings, we select only match-candidate registered voters such that the absolute value of the difference (L1 norm) of their age and the current age is $<=$ 2 years (a blocked age range of 5 years). The intuition for this age bounding leniency is that jail rosters may not consistently update recorded age for booked individuals over time, and the scope of the JDI database is 2 years. For bookings with unknown age, we do not block on voter age.
  \item For each match-candidate booking with the current age value, we select only match-candidate registered voters with equivalent surname soundex codes.
\end{enumerate}

Our sampling process runs over a random selection of within-state match-candidate pairs. We compute a binary match indicator field for parametric initialization by calculating the average of the Jaro-Winkler similarities of first name, last name, and full name, considering matches to be any rows with resultant average $>$ 0.88. We also output numeric similarity vectors for each match-candidate pair. These are defined by the following logic:

\begin{enumerate}
  \item FIPS: 1 if FIPS values are equal; else 0.
  \item Age: 2 if the absolute value of the difference of age between booking and registered voter is $<=$ 1; 1 if either age is missing; else 0.
  \item Gender: 2 if booking and registered voter gender values are equal; 1 if either gender is missing; else 0.
  \item First Name: 2 if Jaro-Winkler similarity of booking and registered voter first name is $>$ 0.94; 1 if Jaro-Winkler similarity of booking and registered voter first name is $>$ 0.88 and $<=$ 0.94; else 0.
  \item Middle Name: 2 if (a) one value is an initial and the other is a complete middle name, and the initial equals the first letter of the complete middle name or (b) both values are complete middle names, and their Jaro-Winkler similarity is $>$ 0.94; 1 if (a) either middle name is missing or (b) both values are complete middle names, and their Jaro-Winkler similarity is $>$ 0.88 and $<=$ 0.94; else 0.
  \item Last Name: 2 if Jaro-Winkler similarity of booking and registered voter last name is $>$ 0.94; 1 if Jaro-Winkler similarity of booking and registered voter last name is $>$ 0.88 and $<=$ 0.94; else 0.
\end{enumerate}

For the synthetic booking and registered voter row examples above, sampling output would appear as follows:

\begin{center}
\begin{tabular}{ |c|c||c|c|c|c|c|c|:c:c:| } 
 \hline
 Booking ID & Registered Voter ID & FIPS & Age & Gender & First & Middle & Last & Average & Match \\ 
 \hline
 12a3b... & LALAL1... & 1 & 2 & 1 & 0 & 2 & 2 & 0.83 & 0 \\
 \hline
\end{tabular}
\end{center}

Sampling results are used to initialize parameters for Fellegi-Sunter calculation. Subsequently, iterative expectation maximization is conducted to stabilize parameter estimates from 50 re-samples of 1 million match-candidate pairs each. Following the full linkage process, we retain only the registered voter match(es) for each booking with the highest probability score(s), and discard all matches with probability scores $<$ 0.5. We then conduct in-block/in-state term frequency re-weighting of first and last names on match records as described by \cite{enamorado} and \cite{winkler}. The final steps of our matching process are exclusions using the following logic:

\begin{enumerate}
    \item We exclude any rows where re-weighted match probability is $<0.75$.
    \item We exclude all rows for any booking that is matched to more than one unique voter ID.
    \item We exclude all rows for any registered voter ID that is matched to overlapping bookings (where booking start and end dates are determined as described in Section \ref{data_jdi} of this report).
    \item We exclude all rows where booking-reported age is $<$ 18.
    \item We exclude all rows where voter registration date is after Election Day (November 3, 2020).
\end{enumerate}

In total, out of our initial sample of 944,985 jail bookings we match to 331,703 registered voter records. As a final step to enable measurement of covariates related to charges against booked individuals, we subset this pool of matches to include only the 305,239 individuals detained in facilities whose rosters report charge information.

\clearpage

\end{onehalfspacing}

\end{document}